\documentclass[11pt]{article}

\usepackage{hyperref}
\usepackage{url} 
\makeatletter
\@ifundefined{doi}{%
	\newcommand{\doi}[1]{%
		\begingroup
		\edef\DOIURL{https://doi.org/\unexpanded{#1}}%
		\href{\DOIURL}{\nolinkurl{\DOIURL}}%
		\endgroup
	}%
}{}
\makeatother

\usepackage[T1]{fontenc}
\usepackage[utf8]{inputenc}
\usepackage{microtype}
\usepackage{graphicx}
\usepackage{booktabs}
\usepackage{siunitx}
\usepackage{amsmath, amssymb}
\usepackage{array}
\usepackage{longtable}
\usepackage{enumitem}
\usepackage{hyperref}
\usepackage{csquotes}
\usepackage{placeins}
\usepackage{subcaption}
\usepackage{tabularx} 
\usepackage{threeparttable} 
\usepackage{ragged2e} 

\usepackage[a4paper,margin=30mm]{geometry}
\usepackage{lmodern}
\usepackage{microtype}
\usepackage{setspace}
\usepackage[numbers,sort&compress]{natbib}
\setcitestyle{square}

\title{A Simulation of Ageing and Care Accessibility in Italian Inner Areas}

\author{
	Roberto Garrone\\[0.5ex]
	\small
	Department of Informatics, Statistics, and Communication,\\
	\small
	University of Milan-Bicocca, DISCo, Milan, Italy\\
	\small
	\texttt{roberto.garrone@unimib.it}
}

\date{\today}

\begin{document}
\maketitle 



\begin{abstract}
Ageing societies face increasing strain on formal and informal care systems, particularly in low-density mountainous municipalities where sparse services and steep terrain constrain access. This study presents a spatially explicit agent-based model that integrates a road-network GIS, synthetic populations derived through Iterative Proportional Fitting, and behavioural heterogeneity to examine how alternative service configurations shape accessibility and caregiver burden. The model, applied to Premeno (Piedmont, Italy), compares a baseline distribution of ambulatory services with a relocation scenario at Villa~Bernocchi. System-level indicators (Caregiver Effort, Overwhelmed Caregivers, Hours~Not~Cared, Walkability) and micro-spatial metrics (Walkability, Detour~Ratio, Proximity) are analysed across 40~batches and 50~stochastic replications per scenario. Results reveal aggregate neutrality but pronounced local redistribution of accessibility. Sensitivity analysis shows that spatial impedance dominates accessibility, whereas behavioural capacity modulates care effort. The findings illustrate distinctive properties of complex adaptive social systems - emergence, heterogeneity, and feedback - demonstrating how computational social simulation can highlight policy trade-offs between spatial efficiency, social equity, and care sustainability in ageing territories.
\end{abstract}





\FloatBarrier
\section{Introduction}
Demographic ageing is transforming the structure and functioning of contemporary societies. As life expectancy rises and family networks shrink, the capacity of public and informal care systems to guarantee accessible and sustainable assistance becomes increasingly uncertain. Peripheral and mountainous areas are particularly exposed: dispersed settlement patterns, steep topography, and scarce public transport combine to amplify inequalities in access to health and social services. Understanding these dynamics requires analytical frameworks able to capture both the spatial heterogeneity of environments and the interactive behaviour of the individuals embedded within them.
Computational social science provides such a framework by combining data-rich representations of social environments with agent-based simulation of micro-level decisions and interactions \citep{cioffi2017computational,conte2014agent}. Within this perspective, agent-based models (ABMs) have been applied to domains ranging from urban mobility and spatial systems \citep{crooks2008agent,heppenstall2016agent} to health behaviour \citep{tracy2018} and ageing policy \citep{auchincloss2019}. Recent work has further situated ABM within population health frameworks \citep{silverman2021}. Yet few studies integrate geographic information systems (GIS), demographic microdata, and survey-derived behavioural attributes in a single, reproducible simulation environment.
This article presents a spatially explicit agent-based model (ABM) that couples a road network GIS with a behaviourally heterogeneous population of elder-caregiver dyads. The model integrates municipal indicators and survey-based parameters via synthetic population techniques to evaluate how ambulatory service reconfiguration affects both \emph{system-level} performance and \emph{micro-spatial} accessibility patterns. The empirical case focuses on the municipality of Premeno (Piedmont, Italy) and compares a baseline configuration (Scenario~1) with a relocation of ambulatory services to Villa Bernocchi (Scenario~2), reflecting a policy option motivated by service rationalisation and cost-containment considerations. The study aims to quantify aggregate differences, reveal sub-group and neighbourhood heterogeneity, and identify policy-relevant trade-offs.

\noindent
\subsection{Positioning and Contribution}
This study advances simulation-based analysis by operationalising ageing-care
accessibility as a \emph{complex adaptive social system}. Despite extensive research
on ageing, accessibility, and service provision, a persistent gap remains in how care
accessibility is conceptualised and analysed in sparsely populated and morphologically
constrained territories. Much of the foundational and applied literature continues to
operationalise accessibility through static indices, optimisation-based location models,
or aggregate planning heuristics that prioritise average distance minimisation or
coverage efficiency at the population level
\citep{hansen1959accessibility,church2000coverage,crooks2008agent,heppenstall2016agent}.
While valuable for descriptive assessment and infrastructure benchmarking, these
approaches typically abstract from individual-level behavioural constraints and
implicitly assume representative users, fixed behavioural responses, and monotonic
relationships between spatial efficiency and welfare outcomes
\citep{geurs2004accessibility,neutens2015accessibility,kwan2012ugc,paez2012accessibility}.

However, ageing-care systems exhibit the defining characteristics of complex
socio-spatial systems, in which mobility limitations, caregiving capacity, and spatial
structure interact non-linearly over time. Empirical and theoretical work in welfare
geography, health services research, and computational social science shows that
interventions improving aggregate accessibility indicators may simultaneously
generate uneven redistributions of burden and opportunity across social groups and
neighbourhoods
\citep{smith2010geography,asada2012equity,conte2014agent,cioffi2017computational}.
Such micro--macro decoupling is difficult to detect using aggregate or optimisation-based
frameworks and remains underexplored in spatially explicit, agent-based analyses of
care accessibility.

To address this gap, the present study unifies three methodological strands:
(i) synthetic population generation through Iterative Proportional Fitting (IPF),
(ii) GIS-embedded agent-based simulation of ageing and caregiving dynamics, and
(iii) global sensitivity analysis for mechanism discovery. Together, these components
form a reproducible and data-rich framework that links the social-data synthesis
tradition of population science with the mechanism-oriented exploration characteristic
of computational social science. By conceptualising accessibility and caregiver burden
as emergent outcomes of decentralised, behaviourally heterogeneous interactions
embedded in a real spatial network \citep{holland1992, gilbert2008agent}, the model
reveals a mechanism that is largely invisible to aggregate or optimisation-based
approaches: \emph{cost-driven spatial reconfiguration can yield aggregate neutrality
	while redistributing accessibility and caregiver burden across neighbourhoods and
	social groups}. The contribution of this work is therefore not a specific policy
prescription, but a demonstration of how micro-level behavioural constraints and
spatial structure decouple efficiency and equity at the system level, illustrating the
analytical value of agent-based simulation for the study of ageing-care systems
\citep{tracy2018, silverman2021, grimm2020}.

\FloatBarrier
\section{Methods}
The empirical case is used as a controlled simulation environment designed to explore system behaviour under alternative service configurations, rather than as a basis for direct policy prescription or empirical forecasting. For clarity, we distinguish between different levels of the simulation design: a \emph{scenario} refers to a structural configuration of the system (e.g., service location layout), a \emph{batch} denotes a specific parameter setting within a scenario, and a \emph{run} corresponds to a single stochastic replicate of a given batch.

\subsection{Model overview}
The model represents a population of dyads composed of older adults and their informal caregivers.  Each dyad operates within a georeferenced environment derived from high-resolution spatial data, including topography, road networks, and service locations. 

\begin{table}[htbp]
	\centering
	\small
	\caption{ODD-D summary for the AWSim model, integrating details from the Supplementary Material on initialization, stochasticity, reproducibility, and submodel design.}
	\label{tab:odd_summary}
	
	\setlength{\tabcolsep}{6pt}
	\renewcommand{\arraystretch}{1.25}
	
	\resizebox{\textwidth}{!}{%
		\begin{minipage}{\textwidth}
			\begin{tabular}{p{0.20\textwidth} p{0.75\textwidth}}
				\toprule
				\textbf{ODD-D Element} & \textbf{Summary of AWSim Implementation} \\
				\midrule
				\textbf{Purpose} &
				To examine how spatial configuration of services, behavioural heterogeneity, and stochastic variability jointly influence accessibility, walkability, and caregiver burden in ageing populations of Italian inner-area municipalities. The model explores trade-offs between elder autonomy, caregiver sustainability, and service coverage efficiency. \\
				
				\textbf{Entities} &
				(1) Elder (patient) agents; (2) Informal caregiver agents; (3) Service facilities (ambulatory, pharmacy, etc.); (4) Road-network nodes representing the GIS environment. \\
				
				\textbf{State variables} &
				For elders: ageing stage, ADL/IADL ability, health severity, care hours required, and mobility radius.  
				For caregivers: available time, income, support network, mobility mode, and effort index.  
				For services: location, capacity, accessibility, and interaction score.  
				Global variables include walkability (WKB), caregiver effort (CEI), hours not cared (HNC), and caregiver overwhelm (COI). \\
				
				\textbf{Temporal scale} &
				Discrete time; 1 tick = 1 hour, with a day subdivided into three sessions (morning, afternoon, evening).  
				Simulations typically run for one month after a 30-iteration warm-up. \\
				
				\textbf{Spatial scale} &
				Municipal-level graph derived from GIS layers (OSM road network, Copernicus DEM).  
				Nodes represent buildings or intersections, mean spacing $\approx$ 25 m.  
				Each node stores elevation, slope, safety, fatigue, and pleasantness scores used in walkability computation. \\
				
				\textbf{Process overview} &
				(1) Update ageing stage and mortality;  
				(2) Recompute caregiver availability and effort;  
				(3) Select service session and compute metrics (e.g., travel time/distance, energy cost)  
				(4) Update functional ability and recompute WKB, CEI, HNC, and COI indices;  
				(5) Record KPIs and aggregate statistics.  
				Time advances synchronously for all agents. \\
				
				\textbf{Design concepts} &
				\emph{Emergence:} accessibility and burden patterns emerge from decentralized dyad decisions.  
				\emph{Adaptation:} agents adjust walking radius, assistance frequency, and mobility mode.  
				\emph{Heterogeneity:} agents differ by age, income, support, and mobility attributes.  
				\emph{Stochasticity:} Gaussian, uniform, and log-normal components govern care frequency, ageing, income, and session duration.  
				\emph{Learning and objectives:} bounded-rational decisions under limited information.  
				\emph{Reproducibility:} ensured through per-agent hashing-based random seeding. \\
				
				\textbf{Input data} &
				Demographic and socioeconomic indicators from ISTAT; GIS data (OSM roads, Copernicus DEM); Structured caregiver interviews; Empirical mobility datasets; Literature-based behavioural parameters. See Supplementary Material~(\S5) \\
				
				\textbf{Initialization} &
				Synthetic population generated via iterative proportional fitting (IPF) and imputation using municipal marginals.  
				Agents are assigned to residential nodes proportional to population density.  
				Services are located according to scenario configuration.  
				Warm-up length determined by Schruben-Singh-Tierney test ($\approx$ 30 ticks).  
				Experiments executed with 40 independent replications and batch-means estimation. \\
				
				\textbf{Submodels} &
				(1) \textit{Walkability Index (WKB)} - slope-, safety-, and pleasantness-weighted accessibility.  
				(2) \textit{Caregiver Effort Index (CEI)} - integrates income, support, and mobility burden.  
				(3) \textit{Hours Not Cared (HNC)} - unmet assistance due to time or income constraints.  
				(4) \textit{Caregiver Overwhelm (COI)} - identifies dyads exceeding effort thresholds.  
				All indices are bounded, continuous, and computed from session-level outcomes. \\
				
				\bottomrule
			\end{tabular}
		\end{minipage}
	}
\end{table}

A simulation run advances in discrete time steps representing days; and, at each step, dyads update their decisions according to local accessibility conditions, accumulated effort, and contextual constraints such as travel distance, slope, and terrain resistance. The model’s primary purpose is to explore how the spatial configuration of essential services and the characteristics of caregivers and elders jointly affect accessibility, caregiving burden, and the overall sustainability of local care systems. 

The model follows the updated ODD-D protocol \citep{grimm2020}, ensuring full transparency and reproducibility. A condensed checklist of its core components is reported in Table~\ref{tab:odd_summary}, while the complete protocol-including purpose, entities, state variables, process scheduling, and design concepts-is provided in the \textit{Supplementary Material}. All stochastic processes (agent initialization, tie assignment, random choice) are controlled by a fixed random seed, ensuring reproducibility across runs. Initialization, warm-up, and randomisation procedures are detailed in the \textit{Supplementary Material}~(\S5-6), including the Schruben-Singh-Tierney test for warm-up length and the hashing-based RNG seeding scheme. The model was developed in NetLogo 6.4 \cite{netlogo} with Python extension for batch execution and sensitivity analysis.

The combination of stage-level and dyad-level attributes enables heterogeneity in both physiological decline and caregiving capacity. Each elder-caregiver pair evolves according to interacting state variables: the elder’s ageing stage and mobility define care demand, while the caregiver’s availability, employment, and support network determine effective supply. Service locations act as environmental constraints linking behavioural rules to spatial structure, allowing system-level accessibility patterns to emerge endogenously from micro-level interactions. These parameterisations collectively ensure that simulated outcomes are consistent with the diversity observed in real-world ageing and caregiving trajectories.

\FloatBarrier
\subsection{Scenarios and experimental design}
Premeno, located in the Verbano-Cusio-Ossola province of Piedmont (Italy), represents a typical mountainous inner-area municipality characterised by low population density and high ageing ratios (Table~\ref{tab:demographics}).

\begin{table}[htbp]
	\centering
	\renewcommand{\arraystretch}{1.1}
	\caption{Demographic context of the case-study municipality compared with the regional and national levels. 
		Premeno exhibits a markedly aged population and high settlement compactness, representative of small mountainous inner-area communities.}
	\label{tab:demographics}
	\begin{tabularx}{\linewidth}{X c c c}
		\toprule
		\textbf{Metric} & \textbf{Premeno} & \textbf{Piedmont} & \textbf{Italy} \\
		\midrule
		Population & 746 & 4.3\,M & 59.4\,M \\
		Urbanized (\%) & 96.4 & 91.1 & 91.0 \\
		Elderly ($>$65\,yr) (\%) & 39.0 & 37.1 & 32.0 \\
		\bottomrule
	\end{tabularx}
\end{table}

Two configurations are analysed for Premeno: \emph{Scenario~1} (baseline distribution of ambulatory services) and \emph{Scenario~2} (relocation to the Villa Bernocchi site). Each scenario was parameterised using empirical and literature-based constants summarised in Table~S4 in the \textit{Supplementary Material}~(\S5.1). Each scenario is executed in 40 experimental batches with 50 independent replications per batch (2{,}000 runs per scenario). Key Performance Indicators (KPIs) include: Caregiver Effort Index (CEI), Caregiver Overwhelmed (CO), Hours Not Cared (HNC), and a Walkability Index (WKB); see Table~\ref{tab:kpi-notation}. Micro-level indices include Walkability (WKB), Effective Detour Ratio (EDR), and Household Proximity Index (HPI). Statistical comparisons employ Welch $t$-tests with multiplicity adjustment where applicable and cluster-robust checks as reported in the scenario analysis. In this study, \emph{accessibility} denotes potential spatial opportunity to reach services, operationalised via the Walkability Index ($\mathrm{WKB}$), whereas \emph{realised care outcomes} are captured by caregiver- and dyad-level indicators, including the Caregiver Effort Index ($\mathrm{CEI}$), Hours Not Cared ($\mathrm{HNC}$), and the Caregiver Overwhelmed indicator ($\mathrm{CO}$).

\begin{figure}[t]
	\centering
	\includegraphics[width=0.80\linewidth]{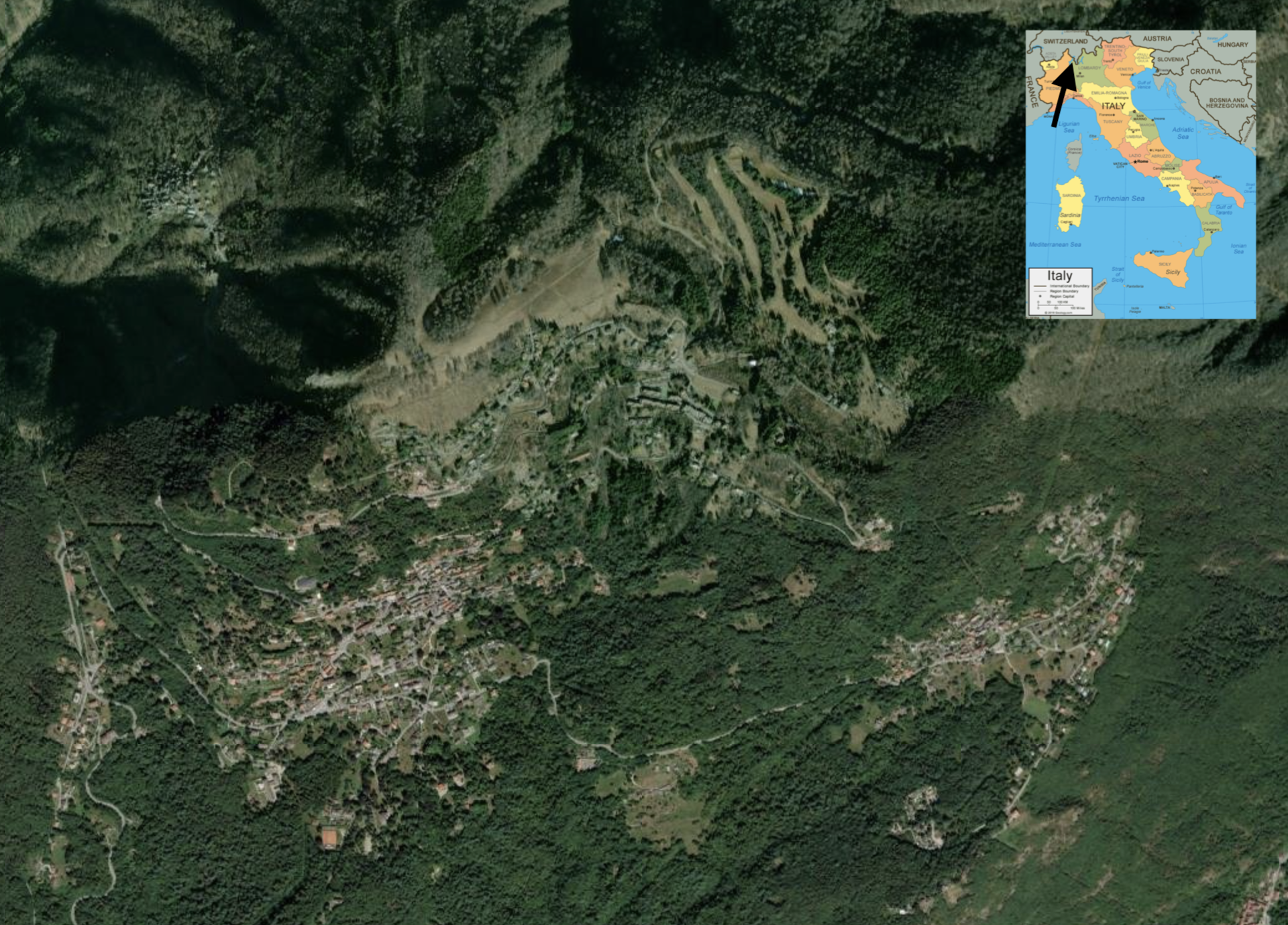}
	\caption{Study area: municipality of Premeno (Piedmont, Italy), an  inner area characterised by low service density and steep morphology. The inset map (top right) shows its geographical position in northern Italy. Source: OpenStreetMap \citep{OSM2024}.}
	\label{fig:premeno_area}
\end{figure}

\FloatBarrier
\subsection{Synthetic population and data integration}
A synthetic population was generated using small-area municipal indicators aligned with national aggregates and survey-derived joint constraints. Iterative Proportional Fitting (IPF) adjusted seed microdata from the ISTAT AVQ 2019 survey until the weighted marginals matched the 8MilaCensus 2011 distributions \citep{Istat2024,istat2022mobility}. Each synthetic record combines demographic, socioeconomic, and health attributes; dyads are then formed according to conditional caregiving probabilities from the AVQ survey. Residential assignment follows population-density weights on the road-network nodes. This procedure yields an empirically grounded yet anonymised micro-dataset reproducing the spatial and social structure of the case-study municipality.

All data used to construct the synthetic population are publicly available from ISTAT, OpenStreetMap \citep{OSM2024}, and the Copernicus Land Monitoring Service \citep{ESA2024}. No individual-level personal data are processed. The synthetic dataset can be regenerated using the procedures described in the supplementary material.

\FloatBarrier
\section{Sensitivity Analysis and Robustness}
A two-stage global sensitivity analysis was conducted following \citet{morris1991factorial} and \citet{saltelli2010variance}. Stage 1 applied the Morris elementary-effects method (10 trajectories, $\Delta=0.5$) for qualitative screening; Stage 2 employed Sobol variance decomposition with Saltelli sampling ($n=512$ samples per parameter, 
$N=2\,000$ total model evaluations). 
Sensitivity analyses were implemented through an automated hybrid pipeline, wherein NetLogo-via its Python extension-triggered a command-line shell to launch a Jupyter notebook executing the \texttt{SALib} routines. This modular configuration enabled seamless coupling between model execution and parameter sampling while enforcing identical random seeds across replications to guarantee methodological reproducibility.

Spatial impedance parameters (e.g., distance-decay and acceptable walking-time thresholds) strongly affected accessibility measures, while parameters related to behavioural capacity and mobility shaped care effort and unmet hours. Replication-based diagnostics showed narrowing confidence intervals with increasing replications and stable rank ordering of parameter influence, supporting numerical robustness (full details in the AWSim technical report referenced below).

\begin{table}[t]
	\centering
	\small
	\setlength{\tabcolsep}{5pt}
	\renewcommand{\arraystretch}{1.15}
	\caption{Key performance indicators (KPIs) and notation used throughout the study. Units refer to the model’s internal time and distance conventions (see Supplementary Material for implementation details).}
	\label{tab:kpi-notation}
	\begin{threeparttable}
		\begin{tabularx}{\linewidth}{@{} l l l >{\RaggedRight\arraybackslash}X >{\RaggedRight\arraybackslash}X @{}}
			\toprule
			\textbf{Symbol} & \textbf{Name} & \textbf{Unit} & \textbf{Interpretation} & \textbf{Aggregation level} \\
			\midrule
			$\mathrm{WKB}$ & Walkability Index & score (0--50) &
			Potential spatial accessibility to services via the road network under mobility constraints &
			node / household / dyad; summarised to municipality \\
			
			$\mathrm{CEI}$ & Caregiver Effort Index & effort score (0--1) &
			Care effort required from the informal caregiver to meet elder needs given accessibility and constraints &
			dyad; summarised by stage / municipality \\
			
			$\mathrm{CO}$ & Caregiver Overwhelmed & number &
			Count of the caregivers unable to provide more care; proxy for overload risk &
			dyad; summarised by stage / municipality \\
			
			$\mathrm{HNC}$ & Hours Not Cared & hours per period &
			Unmet care demand (hours of required care not delivered) under current constraints &
			dyad; summarised by stage / municipality \\
			\bottomrule
		\end{tabularx}
		
		\begin{tablenotes}[flushleft]
			\footnotesize
			\item \emph{Notes.} All KPIs are internally normalized before aggregation. $\mathrm{WKB}$, $\mathrm{CEI}$, and $\mathrm{CO}$ are dimensionless indices derived from time- and network-based quantities via monotone transformations, while $\mathrm{HNC}$ is expressed in physical time units but may be rescaled for comparability. Aggregated values are computed as population- or dyad-weighted means unless stated otherwise. Formal definitions and implementation details for $\mathrm{WKB}$, $\mathrm{CEI}$, $\mathrm{CO}$, and $\mathrm{HNC}$ are provided in the Supplementary Material (SM, Sections S2--S4).
		\end{tablenotes}
	\end{threeparttable}
\end{table}

\begin{table}[t]
	\centering
	\small
	\setlength{\tabcolsep}{5pt}
	\renewcommand{\arraystretch}{1.15}
	\caption{Micro-level spatial and network indices used for mechanism analysis. Indices are computed locally using network-based distances and routing.}
	\label{tab:micro-indices}
	\begin{threeparttable}
		\begin{tabularx}{\linewidth}{@{} l l l >{\RaggedRight\arraybackslash}X >{\RaggedRight\arraybackslash}X @{}}
			\toprule
			\textbf{Symbol} & \textbf{Name} & \textbf{Unit} & \textbf{Interpretation} & \textbf{Aggregation level} \\
			\midrule
			$\mathrm{WKB}$ & Walkability Index & score (0--50) &
			Local potential to reach essential services on foot via the road network under mobility constraints &
			node / household / dyad \\
			
			$\mathrm{EDR}$ & Effective Detour Ratio & ratio ($\geq 1$) &
			Inefficiency of network-based walking routes relative to straight-line distance &
			node / household / dyad \\
			
			$\mathrm{HPI}$ & Household Proximity Index & score (dimensionless) &
			Network-based proximity of a household to nearby households or support nodes &
			household / dyad \\
			\bottomrule
		\end{tabularx}
		
		\begin{tablenotes}[flushleft]
			\footnotesize
			\item \emph{Notes.} All indices use network-based distances; Euclidean distance is used only as a reference for $\mathrm{EDR}$. $\mathrm{WKB}$ values are internally normalised and discretised for reporting. Formal definitions and computational details are provided in the Supplementary Material (SM, Sections S2--S3).
		\end{tablenotes}
	\end{threeparttable}
\end{table}

\subsection{Alternative Methodological Approaches}

Several alternative methodological approaches could, in principle, be used to study
ageing-care accessibility in spatially constrained territories. These approaches are
briefly discussed here to clarify the rationale for adopting an agent-based modelling
framework.

First, static accessibility metrics and GIS-based indicators, such as cumulative
opportunity measures, gravity-based indices, or isochrone analyses, are widely used
to evaluate spatial access to services
\citep{hansen1959accessibility,geurs2004accessibility,paez2012accessibility}.
While effective for benchmarking infrastructure coverage and identifying underserved
areas, these methods treat accessibility as a fixed property of space and population
averages. They do not capture how individual mobility limitations, caregiving capacity,
or scheduling constraints interact dynamically over time, nor how accessibility gains
translate into realised care under heterogeneous behavioural conditions
\citep{kwan2012ugc,neutens2015accessibility}.

Second, optimisation-based location and allocation models, such as maximal coverage,
$p$-median, or network flow formulations, can be employed to design service
configurations that minimise average distance or maximise population coverage under
resource constraints \citep{church2000coverage}. Although valuable for planning and
normative assessment, such models typically assume representative users and
deterministic demand, and they optimise system-level objectives by construction.
Consequently, they are ill-suited to analysing behavioural adaptation, distributional
effects, or the emergence of unintended inequalities following spatial interventions,
particularly in ageing-care systems characterised by heterogeneous constraints
\citep{crooks2008agent,heppenstall2016agent}.

Third, regression-based and quasi-experimental approaches using observational data
have been applied to assess associations between accessibility measures and health or
care outcomes \citep{asada2012equity,smith2010geography}. However, these methods
require strong assumptions about exogeneity and stable unit treatment values, and
they are limited in their ability to explore counterfactual spatial configurations.
This limitation is especially pronounced in small or peripheral municipalities, where
data sparsity and the absence of natural experiments constrain causal inference.

In contrast, the agent-based approach adopted in this study is specifically designed to
address these limitations. By representing accessibility and caregiver burden as
emergent outcomes of decentralised interactions between heterogeneous agents
embedded in a real spatial network, the model enables the exploration of non-linear
dynamics, micro--macro decoupling, and distributional effects that are not accessible
through aggregate, optimisation-based, or purely statistical frameworks
\citep{holland1992,conte2014agent,cioffi2017computational}.
The aim is therefore not to replace existing methods, but to complement them by
providing a mechanism-oriented perspective on ageing-care accessibility in complex
territorial settings, consistent with established principles of agent-based modelling
and computational social science \citep{gilbert2008agent,grimm2020,tracy2018,silverman2021}.

\FloatBarrier
\section{Results and Discussion}

This study contributes to computational social science by demonstrating how ageing-care
accessibility emerges from the interaction between spatial morphology, behavioural
heterogeneity, and decentralised decision-making. Rather than producing uniform
improvements, spatial reconfiguration of services generates compensatory dynamics:
aggregate indicators remain largely unchanged, while accessibility and caregiver burden
are redistributed across neighbourhoods and social groups.

From a theoretical perspective, this result illustrates a key property of complex adaptive
social systems: efficiency-oriented interventions can be system-neutral yet socially
redistributive. The model shows that spatial efficiency and social equity are not
monotonically aligned, as behavioural constraints condition how spatial opportunities
translate into realised accessibility. This mechanism is largely invisible to representative-
agent models or static accessibility metrics, but becomes evident when micro-level
heterogeneity and spatial networks are modelled explicitly.

The study also highlights the importance of multi-scale evaluation. While system-level
indicators suggest stability, micro-spatial and stage-stratified analyses reveal differentiated
effects, underscoring the risk of relying exclusively on aggregate metrics in policy
assessment. Agent-based simulation thus serves not as a forecasting tool, but as an
explanatory device for identifying structural trade-offs and unintended consequences of
policy interventions.

\FloatBarrier
\subsection{Emergent spatial patterns of accessibility}
A central finding of the simulation is the coexistence of aggregate neutrality with pronounced local and group-level redistribution of accessibility and caregiving burden. While system-level indicators show limited change across scenarios, micro-spatial and stage-stratified metrics reveal differentiated gains and losses, indicating a divergence between macro-level stability and micro-level reallocation. Under Scenario~1, walkability is highly heterogeneous within the municipal boundary. Values in the dedicated walkability analysis range approximately between 12 and 47 with a mean near 27, consistent with fragmented connectivity on steep terrain and distance to services. Under Scenario~2, the mean Walkability Index increases to about 34 and the variance decreases, indicating improved overall accessibility and greater spatial uniformity. However, micro-level indicators reveal a trade-off: route indirectness (EDR) increases and proximity (HPI) decreases for some central households. The relocation thus yields a more walkable environment with slightly less direct access for particular neighbourhoods (see Table~\ref{tab:stage_microindices}). Such micro--macro divergence is a characteristic feature of agent-based representations of social systems, where aggregate stability can coexist with substantial local redistribution driven by heterogeneous constraints and interactions \citep{conte2014agent, cioffi2017computational}.

\begin{table}[htbp]
	\centering
	\renewcommand{\arraystretch}{1.15}
	\caption{Stage-stratified effects on micro-indices by scenario. Scenario~2 increases the detour ratio (EDR) consistently; walkability (WKB) gains are uniform and robust; proximity (HPI) declines in some stages. Means are per stage; $p$-values are from stage-wise tests reported in the scenario analysis.}
	\label{tab:stage_microindices}
	\begin{tabularx}{\linewidth}{l c c c c c X}
		\toprule
		\textbf{Stage} & \textbf{Index} & \textbf{S1 mean} & \textbf{S2 mean} & $\boldsymbol{\Delta}$ \textbf{(S2-S1)} & \textbf{$p$-value} & \textbf{Direction} \\
		\midrule
		1 & EDR & 1.8303 & 2.7419 & +0.9116 & 0.000279 & S2 $>$ S1 \\
		& HPI & 4.1958 & 4.3000 & +0.1042 & 0.753918 & n.s. \\
		& WKB & 34.3289 & 34.8505 & +0.5216 & $<\!10^{-6}$ & S2 $>$ S1 \\[3pt]
		
		2 & EDR & 1.8284 & 2.9466 & +1.1182 & 0.000012 & S2 $>$ S1 \\
		& HPI & 4.8708 & 4.0875 & $-0.7833$ & 0.006894 & S2 $<$ S1 \\
		& WKB & 32.3078 & 32.8436 & +0.5358 & $<\!10^{-6}$ & S2 $>$ S1 \\[3pt]
		
		3 & EDR & 1.7310 & 2.8449 & +1.1139 & 0.000023 & S2 $>$ S1 \\
		& HPI & 4.3833 & 4.0875 & $-0.2958$ & 0.430495 & n.s. \\
		& WKB & 31.3211 & 31.8530 & +0.5319 & $<\!10^{-6}$ & S2 $>$ S1 \\[3pt]
		
		4 & EDR & 1.8688 & 2.8328 & +0.9640 & 0.000099 & S2 $>$ S1 \\
		& HPI & 4.6625 & 4.1000 & $-0.5625$ & 0.059223 & n.s. (trend $\downarrow$) \\
		& WKB & 29.3189 & 29.8289 & +0.5100 & $<\!10^{-6}$ & S2 $>$ S1 \\
		\bottomrule
	\end{tabularx}
\end{table}

\FloatBarrier
\subsection{Behavioural and social heterogeneity}
\noindent
The redistribution observed across scenarios is not solely a spatial effect but is strongly conditioned by behavioural heterogeneity among elder--caregiver dyads. Differences in mobility, caregiving capacity, income, and support networks shape how spatial accessibility translates into realised care outcomes. Behavioural heterogeneity mediates how spatial change translates into caregiving burden. Higher-mobility dyads maintain acceptable accessibility even in peripheral areas, while mobility-limited dyads experience accelerated burden once care-capacity thresholds are reached. In the Premeno experiment, clusters of high-burden dyads coincide with lower walkability and higher dependency, evidencing differentiated outcomes across ageing stages. These patterns illustrate how individual-level variation and local morphology interact to produce uneven but interpretable social outcomes.  
Stage-wise results (Table~\ref{tab:stage_microindices}) show that walkability (WKB) improves consistently across all stages, while the effective detour ratio (EDR) rises by roughly +0.9 to +1.1, signalling a spatial trade-off between usability and route efficiency. Proximity (HPI) exhibits small and often non-significant declines, reflecting marginal losses for central households. These findings reinforce the role of behavioural heterogeneity as a conditioning mechanism that filters spatial opportunity, a dynamic that cannot be represented by representative-agent or purely spatial accessibility models \citep{gilbert2008agent, tracy2018}.

\FloatBarrier
\subsection{System-level and micro-level effects}
\noindent
The relocation scenario considered in this study reflects a cost-containment and service-rationalisation logic frequently discussed in small-municipality health planning. Rather than assuming accessibility improvement as an objective, the scenario provides a test case for examining the distributive consequences of efficiency-driven spatial reconfiguration. System-level KPI comparisons between Scenario~1 and Scenario~2 do not yield statistically significant differences after correction: caregiver effort changes are small, caregiver overwhelm declines marginally, walkability changes are negligible at the aggregate level, and hours not cared display a modest increase when using cluster-robust checks as reported in the scenario analysis (Table~\ref{tab:system_kpis}).  
By contrast, micro-spatial indicators reveal clear redistribution of accessibility: Scenario 2 enhances walkability and spatial uniformity while slightly increasing route detours (Table~\ref{tab:stage_microindices}). From this perspective, the simulation highlights how efficiency-oriented interventions may reallocate disadvantage without producing aggregate welfare gains, a pattern consistent with broader discussions of resilience and unintended effects in public health and welfare systems \citep{douglas2020, ziglio2017}.

\begin{table}[htbp]
	\centering
	\renewcommand{\arraystretch}{1.15}
	\caption{System-level KPIs (global means across replications) by scenario. 
		Aggregate differences are small and not significant after multiplicity adjustment.}
	\label{tab:system_kpis}
	\begin{tabularx}{\linewidth}{l c c c X}
		\toprule
		\textbf{KPI} & \textbf{Scenario 1} & \textbf{Scenario 2} & $\boldsymbol{\Delta}$ \textbf{(S2-S1)} & \textbf{Significance} \\
		\midrule
		Caregiver Effort & 0.16 & 0.17 & +0.01 & n.s. (Holm) \\
		Hours Not Cared  & 1.86 & 2.16 & +0.30 & n.s. (Holm) \\
		Caregiver Overwhelmed & -- & -- & -- & n.s. (see text) \\
		Walkability (aggregate) & -- & -- & $\approx 0$ & n.s.; see micro-indices \\
		\bottomrule
	\end{tabularx}
\end{table}

\noindent
As with all agent-based models, the present simulation is intended as an explanatory and exploratory device rather than a predictive instrument. Its scope is bounded by stylised behavioural rules, exogenously specified institutional configurations, and simplified representations of health and care trajectories. This divergence between macro-neutrality and micro-variation illustrates that spatial interventions reallocate relative advantage rather than uniformly improving welfare. Such heterogeneous effects, typical of complex adaptive systems, highlight the need for multi-scale evaluation when interpreting policy outcomes. Explicitly acknowledging these boundary conditions clarifies the class of questions the model is designed to address and aligns with established best practices for transparent and reproducible agent-based modelling \citep{grimm2020, conte2014agent}. Accordingly, validation is approached in an exploratory and structural sense, focusing on the internal coherence of mechanisms, the plausibility of agent behaviours, and the robustness of observed patterns under parameter variation. Rather than seeking point-wise empirical prediction, the model is intended to assess whether hypothesised interactions between spatial structure, behavioural constraints, and service configuration are sufficient to generate the observed qualitative patterns of accessibility and caregiver burden. Global sensitivity analysis is therefore used as a primary validation instrument to identify influential parameters and to distinguish structural effects from artefacts of parameterisation. This approach is consistent with established practices in agent-based and simulation modelling when the aim is to explore mechanisms and trade-offs in complex social systems.

\FloatBarrier
\begin{figure}[!t]
	\centering
	\begin{subfigure}{1\linewidth} 
		\centering
		\includegraphics[width=\linewidth]{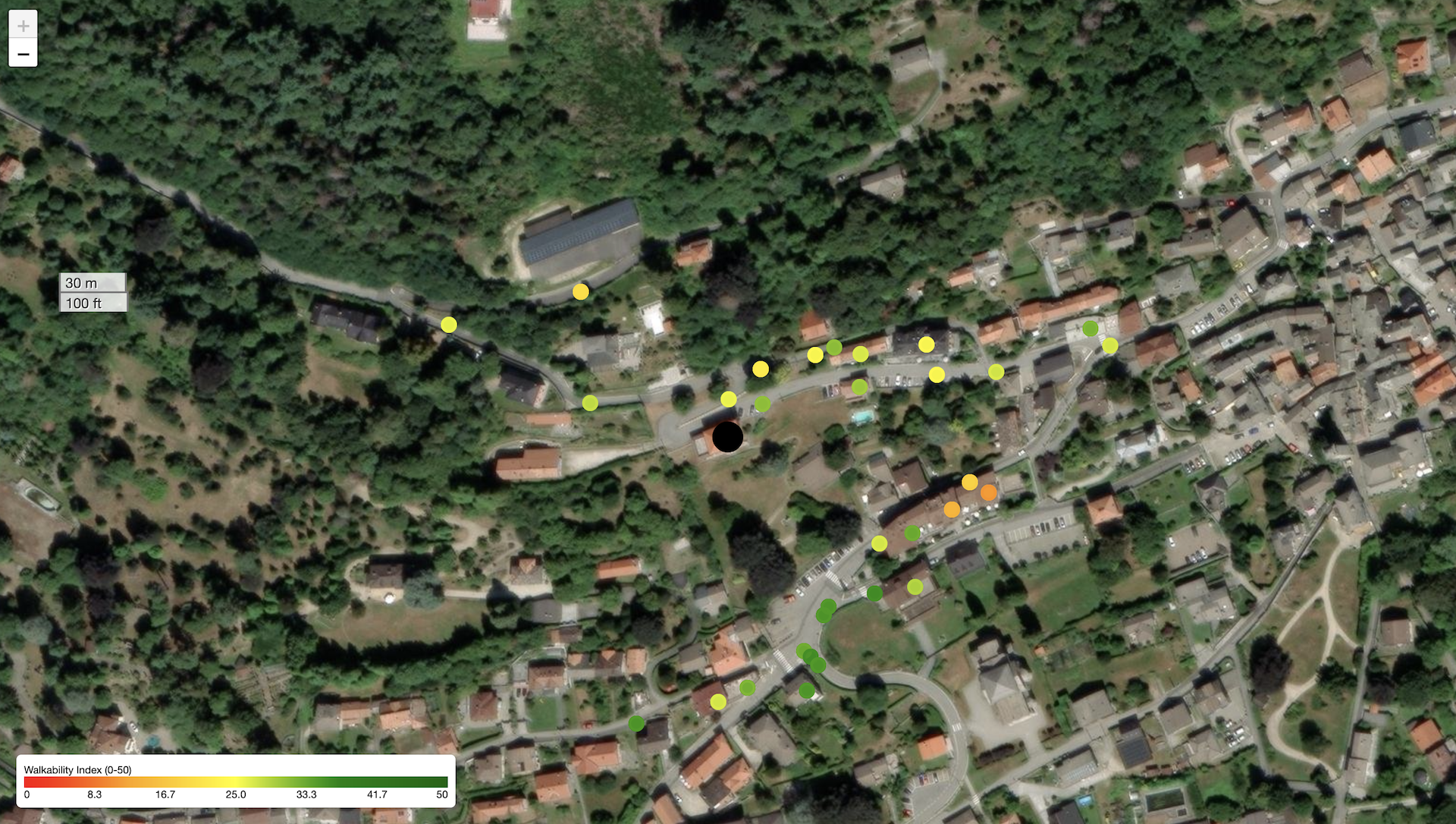}
		\caption*{}
	\end{subfigure}
	\begin{subfigure}{1\linewidth}
		\centering
		\includegraphics[width=\linewidth]{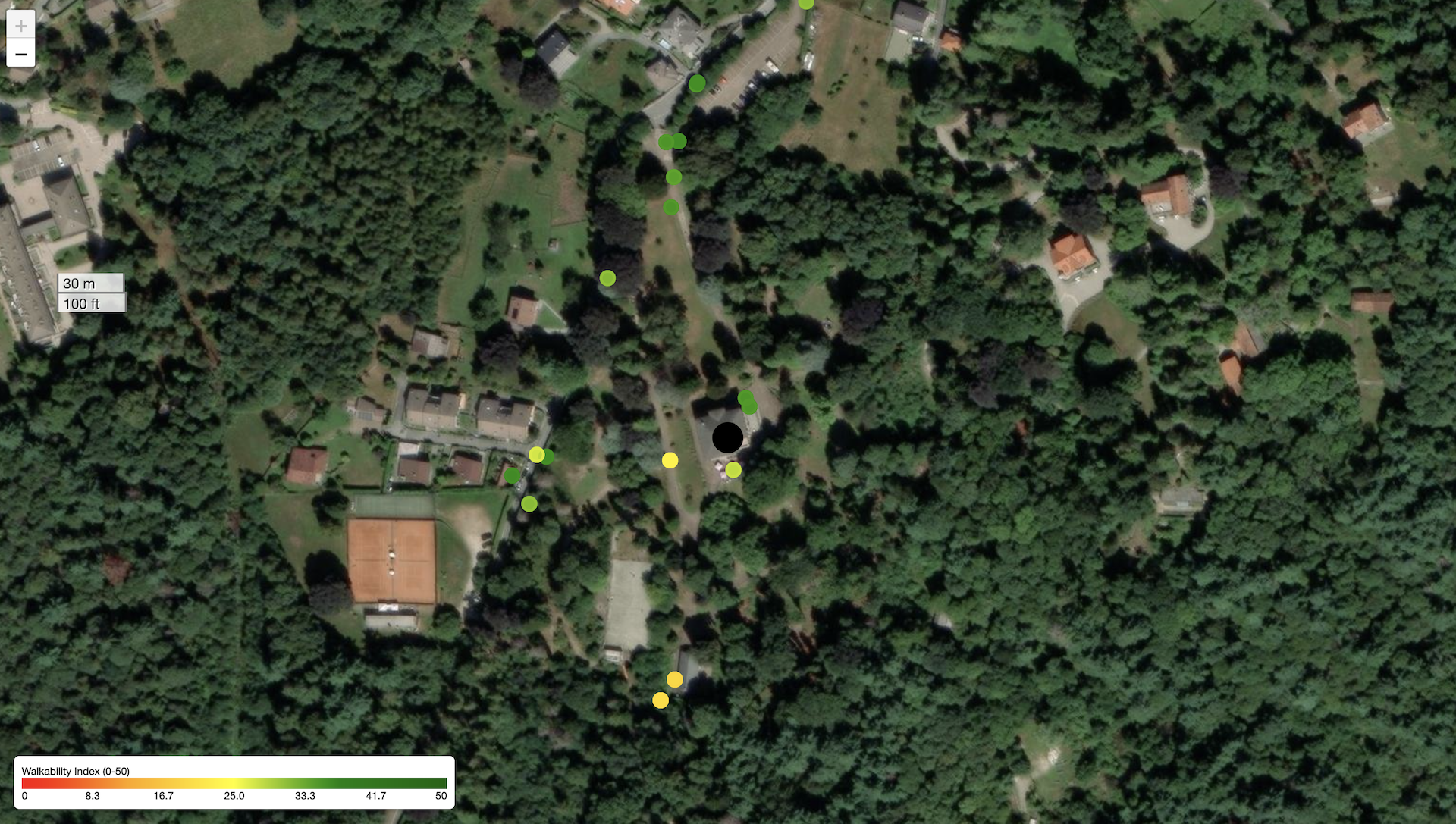}
	\end{subfigure}
\caption{Comparison of ambulatory service locations in the two simulated configurations.
	\emph{Top:} Scenario~1 (current distribution of ambulatory services) shows fragmented accessibility with Walkability Index values ranging from 12 to 47 (mean $\approx$ 27).
	\emph{Bottom:} Scenario~2 (relocation to Villa Bernocchi) exhibits a more homogeneous spatial distribution and higher average walkability values (mean $\approx$ 34), relative to Scenario~1, while maintaining service coverage for central areas.
	Grey-scale points represent residential nodes weighted by Walkability Index (0--50). Values shown are node-level WKB scores, aggregated over runs as means.
	Source: OpenStreetMap \citep{OSM2024}, simulation data.}

	\label{fig:service_scenarios}
\end{figure}

\subsection{Sensitivity-driven insights}
\noindent
Beyond the specific case study, the results have implications for how ageing-care systems are conceptualised and analysed within simulation-based modelling and analysis. The model demonstrates how spatial morphology, behavioural constraints, and decentralised decision-making jointly shape emergent accessibility and burden patterns. The sensitivity analysis clarifies mechanisms behind the scenario results. Spatial impedance parameters directly govern accessibility outcomes, while behavioural capacity parameters condition the translation of spatial gains into reduced burden. Consequently, spatial redesign alone may not yield measurable welfare improvements without parallel organisational or social-support measures (e.g., additional service days, mobility assistance, tele-care). More generally, this study illustrates how agent-based simulation can be used to formalise structural trade-offs in ageing-care systems by linking micro-level behaviour, spatial structure, and emergent system-level outcomes, consistent with core principles of computational social science and complex adaptive systems theory \citep{cioffi2017computational, holland1992}.

\subsection{Emergent socioeconomic and caregiving regimes}
\label{subsec:clusters}

While aggregate indicators show limited net change between scenarios, the simulation outputs exhibit pronounced heterogeneity at the dyad level. To characterise this heterogeneity, a post-simulation clustering analysis was conducted on aggregated outputs across all experimental batches. The objective is not prediction, but interpretation: to identify recurrent socioeconomic and behavioural regimes through which spatial accessibility and caregiving burden are differentially realised.

Clustering was applied to a standardised feature space combining structural socioeconomic attributes (income, education, employment, job skill, car ownership) and caregiving outcomes (hours of care, caregiver effort, unmet care, and caregiver overwhelm). The number of clusters was selected using concordant internal validity criteria (Silhouette, Davies--Bouldin, and Calinski--Harabasz indices), which consistently supported a four-cluster solution. Alternative clustering formulations (Gaussian Mixture Models and density-based methods) yielded structurally comparable partitions, indicating that the identified regimes are robust to algorithmic choice.

Table~\ref{tab:clusters} summarises the four emergent regimes using cluster-level means of key indicators and qualitative semantic labels derived from their relative positioning with respect to global averages.

\begin{table}[ht]
	\centering
	\caption{Emergent socioeconomic and caregiving regimes identified through clustering of simulation outputs. Labels reflect relative positions with respect to global means of socioeconomic endowment and caregiving indicators.}
	\label{tab:clusters}
	\begin{tabular}{p{1.5cm} p{3.8cm} p{4.8cm} p{4.2cm}}
		\hline
		\textbf{Cluster} & \textbf{Socioeconomic profile} & \textbf{Care profile} & \textbf{Interpretation} \\
		\hline
		C0 & Predominantly lower SES profiles, including a subset of elderly living alone without an active caregiver & 
		Low recorded caregiving effort, negligible care hours, no reported burden &
		Low observed burden arising from a combination of self-managed ageing and limited caregiver activation \\
		C1 & High income, high education, high job skill, high mobility & High effort, long care hours, highest caregiver overwhelm & Resource-rich but overloaded: formalised or intensive care arrangements amplify burden despite favourable SES \\
		C2 & Low income, lower education, limited job skill & High effort, long care hours, moderate burden & Resilient but vulnerable: high time investment compensates for scarce resources, masking future burnout risk \\
		C3 & Mixed SES, gender-imbalanced composition & High perceived effort, short care hours, low overwhelm & Emotionally intensive care: low time commitment but high subjective strain \\
		\hline
	\end{tabular}
\end{table}

Several insights follow. First, caregiver burden is not monotonically associated with socioeconomic disadvantage. The cluster exhibiting the highest overload is socioeconomically advantaged (C1), suggesting that resource availability may enable more demanding or formalised care trajectories that intensify time pressure and psychological strain. Second, low observed burden does not necessarily imply favourable conditions: clusters C0 and C3 combine low overload with indicators of latent constraint or emotional stress, which would be invisible in aggregate statistics. Third, the coexistence of C1 and C2 highlights a key decoupling between time investment and perceived burden, reinforcing the interpretation of caregiving as a multidimensional process shaped by both material resources and behavioural adaptation. The interpretation of C0 requires particular care. This cluster includes a heterogeneous set of agents, among which elderly individuals living alone and self-managing daily needs are represented alongside dyads with minimal caregiving activation. Low observed burden in this regime therefore does not necessarily indicate favourable accessibility conditions, but rather reflects limited engagement with caregiving processes, either because needs remain below intervention thresholds or because care is managed autonomously. As a result, C0 remains weakly coupled to both spatial accessibility improvements and caregiving redistribution mechanisms. This distinguishes it structurally from caregiver-mediated regimes (C1--C3), which respond more directly to changes in service configuration.

These regimes provide a behavioural interpretation layer for the scenario results. Aggregate neutrality between service configurations conceals redistribution across clusters: spatial interventions reallocate accessibility and effort unevenly across socioeconomic profiles rather than improving outcomes uniformly. In this sense, clustering complements the micro-spatial analysis by identifying \emph{who} gains or loses under structural change, not merely \emph{where} changes occur. This reinforces the interpretation of the ageing-care system as a complex adaptive system in which efficiency-oriented spatial reconfiguration redistributes burden across heterogeneous agent groups without necessarily improving aggregate welfare.

\FloatBarrier
\subsection{Model robustness and theoretical implications}
These results resonate with broader theories of social resilience and welfare ecology, which frame care systems as adaptive socio-technical networks balancing efficiency and equity under resource constraints \citep{douglas2020,ziglio2017,Keck2013}. In this perspective, the redistribution of accessibility observed in the simulation mirrors real-world adaptive cycles-where structural interventions shift local equilibria without necessarily improving aggregate welfare. By explicitly modelling feedbacks among space, behaviour, and resources, the simulation formalises these resilience mechanisms within a computational social-science architecture.

Each scenario comprised forty experimental batches with fifty stochastic replications (2{,}000 total runs), yielding stable outcome distributions with narrow confidence intervals. Rank-order consistency across replications supports numerical robustness, while cluster-robust checks reduce the risk of spurious significance. From a theoretical perspective, the experiment exhibits three hallmark properties of computational social systems: \emph{emergence} (macro accessibility patterns arise from decentralised decisions on a geospatial network), \emph{heterogeneity} (behavioural and demographic diversity produces unequal yet interpretable outcomes), and \emph{feedback} (changes in accessibility redistribute caregiver workload, which in turn modifies spatial demand). These features align with canonical characteristics of complex adaptive systems \citep{holland1992,cioffi2017computational,gilbert2008agent} and illustrate a micro-macro bridge consistent with computational social science.

\subsection{Limitations and Future Work} 
Several limitations delimit the scope of the present work. Behavioural rules are
stylised, institutional dynamics are exogenously specified, and health trajectories are
simplified. The model does not capture endogenous policy adaptation, learning by
institutions, or long-term demographic feedbacks. Accordingly, results should be
interpreted as structural insights rather than empirical predictions.

Future research may extend the framework in three directions. First, incorporating
adaptive policy agents would allow exploration of endogenous governance dynamics.
Second, coupling the model with longitudinal empirical data could support calibration
of behavioural rules and validation of emergent patterns. Third, comparative applications
across multiple municipalities would enable systematic investigation of how morphology,
demography, and service density jointly shape accessibility regimes.

\FloatBarrier
\section{Conclusions and Policy Implications}
By making explicit the mechanisms through which spatial and behavioural factors interact,
the study demonstrates how agent-based modelling can inform ageing and care policy
by revealing not only whether interventions work on average, but for whom and where
their consequences are most pronounced. Service relocation in Premeno produces aggregate neutrality with local redistribution: 
system-level indicators remain statistically unchanged, whereas micro-spatial metrics reveal 
differentiated gains and losses. Behavioural heterogeneity mediates these effects, 
indicating that infrastructure redesign must be complemented by organisational 
and social-support interventions to achieve equitable burden reduction. 
Beyond the case study, the framework is transferable to other 
European small-municipality contexts where ageing, morphology, and service distribution 
interact. Its modular architecture enables comparative analyses and scenario testing 
for evidence-based policy in health-access planning.

\FloatBarrier
\section*{Acknowledgments}
The authors thank the municipal stakeholders for facilitating access to spatial information 
and providing assistance during site visits and documentation studies.

\FloatBarrier
\section*{Funding and Compliance Statement}
This research was conducted within the \textsc{AGE-IT} project, Spoke~5  ``Care Sustainability in an Ageing Society,'' co-funded by the European Union - NextGenerationEU 
 through the Italian National Recovery and Resilience Plan (PNRR), 
 Mission~4, Component~2, Investment~1.3, under the Extended Partnership~PE8 
 \textit{``Conseguenze e sfide dell’invecchiamento''} 
 (Project \textsc{Age-It} - \textit{A Novel Public-Private Alliance to Generate Socioeconomic, Biomedical and Technological Solutions for an Inclusive Italian Ageing Society - Ageing Well in an Ageing Society}; 
 Grant~Agreement~No.~AGE-IT-PE00000015; CUP:~H43C22000840006).

The simulator employed in this study is used exclusively for research and experimental purposes 
in collaboration with municipal authorities. Data use complies with the General Data Protection Regulation 
(Reg.~(EU)~2016/679) and with the licensing terms of the Italian National Institute of Statistics (ISTAT). 
The system falls under the research exemption of the EU Artificial Intelligence Act 
(Reg.~(EU)~2024/1689) and is not classified as medical device software under 
Reg.~(EU)~2017/745.

\vspace{0.5em}
\noindent\textbf{Disclaimer.} 
The views and opinions expressed in this article are those of the author alone and do not necessarily reflect those of the AGE-IT consortium, its partners, or the funding authority. 

\FloatBarrier
\section*{Declarations}
\textbf{Ethics approval:} Not applicable.\\
\textbf{Consent to participate / publish:} Not applicable.\\
\textbf{Data availability:} Simulation data and code are available from the corresponding author upon reasonable request. Public datasets (municipal indicators and spatial layers) are referenced in the technical documentation.\\
\textbf{Competing interests:} The authors declare no competing interests.\\
\textbf{AI:} During the preparation of this work, the author used ChatGPT (OpenAI) to assist with language editing, clarity improvement, and formatting of the manuscript. After using this tool, the author reviewed and edited the content as needed and takes full responsibility for the content of the published article.\\
\textbf{ORCID:} Garrone: \href{https://orcid.org/0009-0005-7060-6774}{0009-0005-7060-6774}.

\FloatBarrier
\section*{Supplementary Material}
The Supplementary Material provides a complete technical companion to the AWSim model, offering extended methodological and computational details that complement the main manuscript. It follows the updated ODD-D protocol \citep{grimm2020}, ensuring transparency and reproducibility in agent-based model reporting. Specifically, the Supplementary Material includes:

\begin{enumerate}[label=(\alph*)]
	\item \textbf{Full ODD-D description:} A comprehensive account of the model’s purpose, entities, state variables, spatial and temporal structures, process scheduling, and design concepts. The description follows the ODD-D formalism and integrates tables summarizing key attributes of elder, caregiver, and service agents, as well as schematic representations of submodels and data flows.
	
	\item \textbf{Stochasticity and randomization methods:} Detailed discussion of random components affecting dyad selection, aging transitions, walking preferences, assistance hours, and income variability. The document specifies the use of NetLogo’s Mersenne Twister RNG, per-agent substreams, and a hashing-based seeding scheme for reproducibility.
	
	\item \textbf{Initialization, warm-up, and run design:} A dedicated section outlines model initialization from synthetic populations generated via Iterative Proportional Fitting (IPF), warm-up length estimation using the Schruben-Singh-Tierney test with batch means, and replication design for statistical convergence.
	
	\item \textbf{Parameterisation tables:} Complete listings of baseline and empirical parameters (e.g., mobility speeds, income levels, effort thresholds, and ageing transition rates), each with corresponding data sources (ISTAT, survey data, literature benchmarks). A summary of initialization parameters is provided in Table 5.
	
	\item \textbf{Sensitivity and uncertainty analysis:} Extended results from global sensitivity analyses using Morris and Sobol-Saltelli methods, illustrating first-order and total-order effects of financial and mobility parameters on key indicators such as Walkability (WKB) and Caregiver Effort (CEI).
	
	\item \textbf{Scenario results and maps:} Comparative figures showing baseline (\textit{Premeno}) and relocation (\textit{Villa Bernocchi}) scenarios, including heatmaps of walkability, service reachability, and caregiver burden across spatial and temporal scales.
\end{enumerate}

All supplementary files-including the full ODD-D protocol, parameter tables, initialization datasets, and simulation code-are available from the corresponding author upon reasonable request. The documentation ensures that every modeling decision, random process, and calibration parameter can be independently verified and replicated.


\begin{thebibliography}{99}
		
		\bibitem[{Asada(2012)}]{asada2012equity}
		Asada, Y. (2012).
		\newblock A framework for measuring health inequity.
		\newblock \textit{Journal of Epidemiology and Community Health},
		\textit{66}(6), 494--499.
		\newblock \doi{10.1136/jech-2011-200016}
		
		\bibitem[{Auchincloss et~al.(2019)}]{auchincloss2019}
		Auchincloss, A.~H. et~al. (2019).
		\newblock Agent-based models of health behaviour: insights and challenges.
		\newblock \textit{Annual Review of Public Health}, \textit{40}, 1--23.
		\newblock \doi{10.1146/annurev-publhealth-040617-013631}
		
		\bibitem[{Church \& ReVelle(2000)}]{church2000coverage}
		Church, R. \& ReVelle, C. (2000).
		\newblock The maximal covering location problem.
		\newblock \textit{Papers in Regional Science}, \textit{79}(2), 149--165.
		\newblock \doi{10.1007/s101100050042}
		
		\bibitem[{Cioffi-Revilla(2017)}]{cioffi2017computational}
		Cioffi-Revilla, C. (2017).
		\newblock \textit{Introduction to Computational Social Science: Principles and
			Applications}.
		\newblock Cham: Springer, 2 edn.
		\newblock \doi{10.1007/978-3-319-50131-4}
		
		\bibitem[{Conte \& Paolucci(2014)}]{conte2014agent}
		Conte, R. \& Paolucci, M. (2014).
		\newblock On agent-based modeling and computational social science.
		\newblock \textit{Frontiers in Psychology}, \textit{5}, 668
		
		\bibitem[{Crooks et~al.(2008)Crooks, Castle \& Batty}]{crooks2008agent}
		Crooks, A.~T., Castle, C. J.~E. \& Batty, M. (2008).
		\newblock Agent-based modelling for the science of cities.
		\newblock \textit{Computers, Environment and Urban Systems}, \textit{32}(6),
		417--430
		
		\bibitem[{Douglas et~al.(2020)Douglas, Katikireddi, Taulbut, McKee \&
			McCartney}]{douglas2020}
		Douglas, M., Katikireddi, S.~V., Taulbut, M., McKee, M. \& McCartney, G.
		(2020).
		\newblock Mitigating the wider health effects of covid-19 pandemic response.
		\newblock \textit{BMJ}, \textit{369}, m1557
		
		\bibitem[{{European Space Agency}(2024)}]{ESA2024}
		{European Space Agency} (2024).
		\newblock Copernicus global digital elevation model.
		\newblock Distributed by OpenTopography.
		\newblock Accessed: 2024-11-08
		
		\bibitem[{Geurs \& van Wee(2004)}]{geurs2004accessibility}
		Geurs, K.~T. \& van Wee, B. (2004).
		\newblock Accessibility evaluation of land-use and transport strategies: review
		and research directions.
		\newblock \textit{Journal of Transport Geography}, \textit{12}(2), 127--140.
		\newblock \doi{10.1016/j.jtrangeo.2003.10.005}
		
		\bibitem[{Gilbert(2008)}]{gilbert2008agent}
		Gilbert, N. (2008).
		\newblock \textit{Agent-Based Models}.
		\newblock London: SAGE Publications
		
		\bibitem[{Grimm et~al.(2020)Grimm, Railsback, Vincenot et~al.}]{grimm2020}
		Grimm, V., Railsback, S.~F., Vincenot, C.~E. et~al. (2020).
		\newblock The odd protocol for describing agent-based and other simulation
		models: A second update to improve clarity, replication, and structural
		realism.
		\newblock \textit{Journal of Artificial Societies and Social Simulation},
		\textit{23}(2), 7.
		\newblock \doi{10.18564/jasss.4259}
		
		\bibitem[{Hansen(1959)}]{hansen1959accessibility}
		Hansen, W.~G. (1959).
		\newblock How accessibility shapes land use.
		\newblock \textit{Journal of the American Institute of Planners},
		\textit{25}(2), 73--76.
		\newblock \doi{10.1080/01944365908978307}
		
		\bibitem[{Heppenstall et~al.(2016)Heppenstall, Crooks, See \&
			Batty}]{heppenstall2016agent}
		Heppenstall, A., Crooks, A., See, L. \& Batty, M. (2016).
		\newblock \textit{Agent-Based Models of Geographical Systems}.
		\newblock Dordrecht: Springer.
		\newblock \doi{10.1007/978-90-481-8927-4}
		
		\bibitem[{Holland(1992)}]{holland1992}
		Holland, J.~H. (1992).
		\newblock Complex adaptive systems.
		\newblock \textit{Daedalus}, \textit{121}(1), 17--30
		
		\bibitem[{{ISTAT}(2023)}]{istat2022mobility}
		{ISTAT} (2023).
		\newblock Multiscopo sulle famiglie: Aspetti della vita quotidiana --
		mobilit{\`a}.
		\newblock Istituto Nazionale di Statistica, Rome
		
		\bibitem[{{ISTAT}(2024)}]{Istat2024}
		{ISTAT} (2024).
		\newblock 8mila census.
		\newblock Italian National Institute of Statistics.
		\newblock Accessed: 2024-12-12
		
		\bibitem[{Keck \& Sakdapolrak(2013)}]{Keck2013}
		Keck, M. \& Sakdapolrak, P. (2013).
		\newblock What is social resilience? lessons learned and ways forward.
		\newblock \textit{Erdkunde}, \textit{67}(1), 5--19.
		\newblock \doi{10.3112/erdkunde.2013.01.02}
		
		\bibitem[{Kwan(2012)}]{kwan2012ugc}
		Kwan, M.-P. (2012).
		\newblock {The Uncertain Geographic Context Problem}.
		\newblock \textit{Ann. Assoc. Am. Geogr.}
		
		\bibitem[{Morris(1991)}]{morris1991factorial}
		Morris, M.~D. (1991).
		\newblock Factorial sampling plans for preliminary computational experiments.
		\newblock \textit{Technometrics}, \textit{33}(2), 161--174.
		\newblock \doi{10.1080/00401706.1991.10484804}
		
		\bibitem[{Neutens(2015)}]{neutens2015accessibility}
		Neutens, T. (2015).
		\newblock Accessibility, equity and health care: review and research directions
		for transport geographers.
		\newblock \textit{Journal of Transport Geography}, \textit{43}, 14--27.
		\newblock \doi{10.1016/j.jtrangeo.2014.12.006}
		
		\bibitem[{{OpenStreetMap contributors}(2024)}]{OSM2024}
		{OpenStreetMap contributors} (2024).
		\newblock Openstreetmap data.
		\newblock Distributed by OpenStreetMap Foundation.
		\newblock Accessed: 2024-10-03
		
		\bibitem[{P{\ifmmode\acute{a}\else\'{a}\fi}ez(2012)}]{paez2012accessibility}
		P{\ifmmode\acute{a}\else\'{a}\fi}ez, A. {\&}.~S. (2012).
		\newblock {Measuring accessibility: positive and normative implementations of
			various accessibility indicators}.
		\newblock \textit{Journal of Transport Geography}, \textit{25}(C), 141--153
		
		\bibitem[{Saltelli \& Annoni(2010)}]{saltelli2010variance}
		Saltelli, A. \& Annoni, P. (2010).
		\newblock Variance based sensitivity analysis of model output: Design and
		estimator for the total sensitivity index.
		\newblock \textit{Computer Physics Communications}, \textit{181}(2), 259--270
		
		\bibitem[{Silverman et~al.(2021)Silverman, Gostoli, Picascia, Almagor, McCann,
			Shaw \& Angione}]{silverman2021}
		Silverman, E., Gostoli, U., Picascia, S., Almagor, J., McCann, M., Shaw, R. \&
		Angione, C. (2021).
		\newblock Situating agent-based modelling in population health research.
		\newblock \textit{Emerging Themes in Epidemiology}, \textit{18}, 12
		
		\bibitem[{Smith \& Easterlow(2010)}]{smith2010geography}
		Smith, S.~J. \& Easterlow, D. (2010).
		\newblock \textit{The Geography of Health Inequalities}.
		\newblock Oxford: Blackwell
		
		\bibitem[{Tracy et~al.(2018)Tracy, Cerd{\'a} \& Keyes}]{tracy2018}
		Tracy, M., Cerd{\'a}, M. \& Keyes, K.~M. (2018).
		\newblock Agent-based modeling in public health: Current applications and
		future directions.
		\newblock \textit{Annual Review of Public Health}, \textit{39}, 77--94
		
		\bibitem[{Ziglio et~al.(2017)Ziglio, Azzopardi-Muscat \&
			Briguglio}]{ziglio2017}
		Ziglio, E., Azzopardi-Muscat, N. \& Briguglio, L. (2017).
		\newblock Resilience and 21st century public health.
		\newblock \textit{European Journal of Public Health}, \textit{27}(5), 789--790
		
		\bibitem[{Wilensky(1999)}]{netlogo}
		Wilensky, U. (1999).
		\newblock \textit{NetLogo} (software).
		\newblock Center for Connected Learning and Computer-Based Modeling,
		Northwestern University, Evanston, IL.
		\newblock \url{http://ccl.northwestern.edu/netlogo/}
		

	\end{thebibliography}
\end{document}


\setcounter{section}{0}
	\setcounter{table}{0}
	\setcounter{figure}{0}
	\setcounter{equation}{0} 
    \setcounter{table}{0}

	\maketitle
	
	\section{Model scope and conceptual framing}
	
	During early development, it was recognized that fully replicating the complexities of care systems would be infeasible. Accordingly, the modeling strategy focused on capturing core structural and behavioral challenges rather than reproducing the entire institutional architecture, following established modeling guidance \citep{northridge2016}. 
	
	The model centers on the interaction between public healthcare services-represented as facilities-and the demand generated by dyads of patients and informal caregivers. It examines the competing objectives of three stakeholder groups: patients seeking timely access, caregivers constrained by time and income, and service providers aiming to allocate working hours efficiently. This triadic tension embodies the broader trade-off between system efficiency and patient-centered accessibility. Policymakers are not modeled as agents but influence outcomes indirectly through structural parameters such as facility capacity, location, and support availability.
	
	At the micro level, two agent classes-patients ($p \in P$) and informal caregivers ($ic \in IC$)-simulate individual decision-making and interdependence. Each caregiver is associated with a residential node, age cohort, and demographic and socioeconomic profile defining mobility and available time. Each patient is similarly located, characterized by age, health status, and preferences affecting access mode and timing. Health events may be acute (triggering immediate care-seeking) or chronic (requiring sustained supervision). Caregivers may face potential delays or unexpected expenses. Service access occurs through unscheduled visits to geographically explicit facilities, with decisions shaped by patient condition, preferences, and caregiver constraints.
	
	Each service ($s \in SP$) operates from a fixed location and maintains defined weekly hours. Providers follow operational routines governing scheduling and admissions. Dyad-service interactions generate adaptive feedbacks as both sides adjust to evolving conditions.
	
	The framework adopts several simplifying assumptions that delimit its scope. It models adult populations only, excluding pediatric care and gender-specific differences. Although agents differ by age, health, and ageing stage, decision heuristics for service access are homogeneous. Interactions between comorbid conditions, infectious disease dynamics, and health improvement effects of service use are not represented. Services are homogeneous in duration and unaffected by case complexity or socioeconomic status. Home-based care and multi-tiered provider interactions are omitted, and service availability is assumed continuous, without seasonal or staffing interruptions. The municipality is treated as a closed system, isolated from external health networks.
	
	\section{ODD-D Protocol}
	This section details the model following the updated ODD-D framework \citep{grimm2020b}, which standardizes the documentation of agent-based and data-driven simulation models. The framework provides a structured representation of the model’s purpose, entities, processes, scales, and design concepts, ensuring transparency, reproducibility, and comparability across studies.
	To support concise communication of these components, Table 1 at page 3 of the article summarizes the main ODD-D elements of the AWSim model, linking each structural and behavioral component to the corresponding implementation described in the supplementary material.

	\subsection{Purpose}
	The model investigates how service location and behavioural heterogeneity affect accessibility and caregiver burden in ageing societies, focusing on Italian inner areas. It focuses on the interaction among three stakeholder groups-patients, informal caregivers, and policymakers-whose objectives are interdependent yet often conflicting. 
	Patients with chronic or acute conditions seek timely treatment from preferred providers, while informal caregivers aim to manage limited time and energy efficiently, avoiding excessive assistance hours. 
	This asymmetry creates a structural tension between continuity of care and the caregiver’s need for sustainable workload scheduling. 
	Policymakers, in turn, must balance service quality with resource constraints, ensuring adequate coverage without imposing unsustainable financial or staffing burdens. 
	Such trade-offs are operationalized through structural parameters such as the number, capacity, and distribution of ambulatory and transport services.
	
	To capture the multidimensional nature of care accessibility, the model integrates two complementary indices. 
	The \textit{Walkability Index} (WKB) quantifies the physical accessibility of the built environment-incorporating slope, elevation, safety, fatigue, and infrastructure adequacy-and reflects the potential for autonomous mobility among older adults. 
	Higher walkability corresponds to greater independence and lower reliance on caregivers for routine activities. 
	The \textit{Caregiver Effort Index} (CEI) measures the burden experienced by informal caregivers as a function of mobility constraints, travel time, and scheduling flexibility, capturing how environmental and organizational factors affect their capacity to provide timely support. 
	By analysing both indices jointly, the model assesses how spatial configuration and workload dynamics interact to determine care sustainability and identify actionable interventions-such as infrastructure improvements, better service coordination, or relocation of care hubs-that can enhance elder autonomy while reducing caregiver strain.

	\subsection{Entities, state variables, and scales}
	Entities include elder (patient) agents, informal caregiver agents, and service nodes (ambulatory facilities).  
	State variables are listed in Tables~\ref{tab:aging}-\ref{tab:patients} (replicated here for completeness).  
	The simulation environment is characterized by both spatial and temporal scales.
	
	\subsubsection{Spatial structure}
	Spatially, the model is defined on a bounded geographic coordinate system 
	\begin{equation}
L := [-x, x] \times [-y, y], \quad x, y \in \mathbb{N},
\end{equation}
	where each location $l \in L$ is geo-referenced by latitude and longitude coordinates.  
	A road network extracted from GIS data is embedded within this spatial domain, supporting explicit agent mobility and interaction.  
	Each node represents a spatial location characterized by topographical and qualitative walkability attributes-such as elevation, slope, stair presence, perceived safety, effort, fatigue, and pleasantness.  
	Each link, corresponding to a street segment, contains information on surface type, pedestrian and cycling accessibility, directionality, and effective distance.  
	Transport infrastructure nodes include additional metadata (e.g., modality type, available capacity, parking availability, and suitability for mobility-impaired individuals).  
	If required, the environment can be extended to incorporate additional dynamics such as traffic flow, air quality, water, or carbon emissions.
	
	\subsubsection{Temporal structure}
	Temporally, the model operates on a continuous timeline segmented into discrete steps and durations.  
	Each temporal element is defined as
	\begin{equation}
t = (\delta, \eta) \in T := \mathbb{N} \times [0, 1),
\end{equation}
	where $\delta$ denotes the calendar day (a non-negative integer) and $\eta \in [0, 14)$ denotes the hour of the day elapsed.  
	A tick-based format allows for unified encoding of both timestamps and durations, with the context clarifying whether a given tick $k$ refers to a point in time or a time interval.  
	For instance, $(38,15) \in T$ corresponds to tick $492$ and lasts $|492 - 734| = 242$ ticks, ending at $(38,16) \in T$ at tick $734$.  
	Each activity performed by a dyad, patient, or informal caregiver is described by both start time and duration in this format.
	
	To replicate real-world scheduling practices in public service, each simulated day is subdivided into morning, afternoon, and evening sessions.  
	A session is defined as
	\begin{equation}
\lambda = (\delta, \gamma) \in \Lambda := \mathbb{N} \times \{0,1,2\},
\end{equation}
	where $\gamma = 0$ denotes the morning, $\gamma = 1$ the afternoon, and $\gamma = 2$ the evening session.  
	To capture weekly recurrences in care schedules, an equivalence relation $\sim$ is introduced on $\Lambda$:
	\begin{equation}
\lambda_i \sim \lambda_j \iff \text{(same weekday and session type)}.
\end{equation}
	By grouping all sessions of equivalent type into equivalence classes $[\lambda]_\sim$, the model captures systematic temporal variations such as weekday-weekend differences or daily operational cycles.
	
	
	\begin{table}[t]
		\centering
		\setlength{\tabcolsep}{8pt}
		\renewcommand{\arraystretch}{1.1}
				\caption{Attributes defining the ageing stages of elder agents. Severity and functional decline parameters control progression and mortality probabilities.}
		\label{tab:aging}
		\begin{tabular}{lll}
			\toprule
			\textbf{Attribute} & \textbf{Domain} & \textbf{Type} \\
			\midrule
			Severity index & $[0,4]$ & Integer \\
			Worsening threshold & $[0,15]$ & Integer \\
			Base assistance & $[0,15]$ & Float (hours) \\
			Worsening in walk & $[0,15]$ & Float \\
			Base ADL ability & $[0,18]$ & Float \\
			Death probability & $[0,1]$ & Float \\
			\bottomrule
		\end{tabular}

	\end{table}
	
	\begin{table}[t]
		\centering
		\setlength{\tabcolsep}{8pt}
		\renewcommand{\arraystretch}{1.1}
				\caption{Attributes of informal caregiver agents. Demographic, socioeconomic, and relational attributes determine available time and support capacity.}
		\label{tab:caregivers}
		\begin{tabular}{lll}
			\toprule
			\textbf{Attribute} & \textbf{Domain} & \textbf{Type} \\
			\midrule
			Age & [15, 65] & Integer (years) \\
			Sex & [M, F] & Categorical \\
			Income & [400-3000] & Float (currency units) \\
			Life expectancy & [20-50] & Integer (years) \\
			Home & Identifier & Node ID \\
			Mobility & [car, public, walk, green] & Encoded integer \\
			Walking radius & [0, 1500] & Integer (metres) \\
			Efforts & Computed & Index \\
			Has-job & [Y, N] & Binary \\
			Skilled-job & [Y, N] & Binary \\
			Single & [Y, N] & Binary \\
			Has-child & [Y, N] & Binary \\
			Education & [Y, N] & Binary \\
			Divorced & [Y, N] & Binary \\
			Supported & [Y, N] & Binary \\
			Hrs-support & [0, 18] & Float (hours) \\
			\bottomrule
		\end{tabular}

	\end{table}
	
	\begin{table}[t]
		\centering
		\setlength{\tabcolsep}{8pt}
		\renewcommand{\arraystretch}{1.1}
				\caption{Attributes of elder (patient) agents. Ageing stage, mobility, and dependency drive the demand for informal and formal care.}
		\label{tab:patients}
		\begin{tabular}{lll}
			\toprule
			\textbf{Attribute} & \textbf{Domain} & \textbf{Type} \\
			\midrule
			Age & [65, 95] & Integer (years) \\
			Sex & [M, F] & Categorical \\
			Income & [400-3000] & Float (currency units) \\
			Life expectancy & [5-25] & Integer (years) \\
			Home & Identifier & Node ID \\
			Mobility & [car, public, walk, green] & Encoded integer \\
			Walking radius & [0, 1500] & Integer (metres) \\
			Aging stage & [0, 4] & Integer (coded) \\
			Hrs-care & [0, 18] & Float (hours) \\
			Has-caregiver & [Y, N] & Binary \\
			Single & [Y, N] & Binary \\
			Has-child & [Y, N] & Binary \\
			Education & [Y, N] & Binary \\
			Divorced & [Y, N] & Binary \\
			\bottomrule
		\end{tabular}

	\end{table}
	
	The model comprises three main agent classes-patients ($p \in P$), informal caregivers ($ic \in IC$), and services ($s \in SP$)-and a set of spatial locations ($l \in L$) defined within a geo-referenced environment.  
	Each agent possesses a combination of static and dynamic attributes representing demographic traits, spatial position, and behavioural states.  
	
	\paragraph{Aging Stages and Health Dynamics.}
	Care activities include support for health-related events-such as chronic illnesses or acute episodes-that cause discomfort and require medical care.  
	The illnesses typical of ageing are modelled within discrete \textit{aging stages}, each representing a severity level and its associated care requirements.  
	An aging stage is formally represented as a tuple:
	\begin{equation}
a = (\sigma, \theta, \alpha, \omega, \psi, \pi)
\end{equation}
	where  
	$\sigma$ denotes severity level (integer),  
	$\theta$ is the threshold for worsening,  
	$\alpha$ is the baseline assistance required (hours),  
	$\omega$ represents the worsening in walking abilities,  
	$\psi$ is the base level of IADL/ADL ability, and  
	$\pi$ denotes the probability of death.
	
	Each aging stage generalizes the traits of age-related conditions through mappings:
	\begin{equation}
f_\theta(\sigma), \quad f_\alpha(\sigma), \quad f_\omega(\sigma), \quad f_\psi(\sigma), \quad f_\pi(\sigma),
\end{equation}
	which yield the expected values for the corresponding variables given a severity $\sigma$.
	For instance, a stage of severity~1 for a patient aged~65 is described by
	$\theta=6$, $\alpha=4.5$, $\omega=12$, $\psi=3.5$, and $\pi=0.05$.
	This indicates a threshold of~6 for worsening to the next stage, an assistance need of~4.5~hours, a~12-unit loss in walking ability, a~3.5/5~functional capacity, and a~5\% probability of death.  
	
	\paragraph{Sessions and Activities.}
	Daily activities are modelled as \textit{sessions}, representing discrete events during which patients perform or receive assistance in activities of daily living (ADL/IADL).  
	Each session is defined as:
	\begin{equation}
s_{t} = (\lambda, ic, p, sp),
\end{equation}
	where $\lambda$ is the session time, $ic$ the attending caregiver (if any), $p$ the patient, and $sp$ the accessed service.  
	While multiple sessions can occur per day for the same patient, only one is active per time step.  
	Sessions encapsulate alternative caregiving strategies, reflecting distinct approaches to coordination between patients and caregivers.  
	
	\paragraph{Service Facilities.}
	Services constitute a network of spatially distributed facilities providing essential territorial support-including pharmacies, food outlets, community offices, outpatient clinics, and emergency care.  
	Each service $sp \in SP$ is defined as:
	\begin{equation}
sp = (l, \delta, \kappa, \chi),
\end{equation}
	where $l \in L$ is its fixed location, $\delta$ is a binary accessibility flag, $\kappa$ is its local capacity, and $\chi$ is an interaction score reflecting potential cross-effects between nearby facilities.  
	
	\paragraph{Informal Caregivers.}
	Each caregiver $ic \in IC$ is associated with a fixed home location $l \in L$, an effort score $\varepsilon_{ic}$, an age category $\tau_{ic}$, and a vector of personal and socioeconomic traits-such as sex, marital and parental status, education, employment, and income-along with preferences regarding mobility, walkable distance, and care type.  
	While location and demographic traits remain static, both income and available time may vary dynamically.  
	Caregivers’ employment type and mobility mode constrain their capacity to assist patients during working hours.  
	
	\paragraph{Patients.}
	Patients represent the primary demand side of the system.  
	Each patient $p \in P$ is characterized by a fixed home location $l \in L$, a current aging stage $a$, an ability score $\psi_p$, an age category $\tau_p$, a set of possible sessions $S_p$, a currently active session $s_t$, and personal traits similar to caregivers (e.g., sex, marital status, education, income).  
	Preferences determine preferred service providers and mobility choices, guiding the selection of sessions.  
	When a new need arises, the patient initiates an assistance-seeking process in collaboration with the assigned caregiver, computes the number of hours of care unmet, and updates the aging stage and care needs based on the ADL/IADL evaluation.

	\subsection{Process Overview and Scheduling}
	
	The model operates as a discrete-time agent-based simulation, a paradigm well suited for representing systems characterized by frequent interactions and collective behaviors evolving over regular timescales.  
	At each time step (\textit{tick}), agents update their states according to behavioral rules and environmental conditions.  
	The uniform temporal structure ensures reproducibility and facilitates the visualization of dynamic processes such as accessibility shifts and burden redistribution.
	
	Formally, time advances in discrete iterations $t \in \mathbb{N}$ until the predefined simulation horizon $T_{max}$ is reached.  
	During each tick, all agents are synchronously evaluated in the following order:
	\begin{enumerate}[label=(\arabic*)]
		\item \textbf{Health and Aging Update.}  
		Each patient reassesses their health status and aging stage, potentially transitioning to a higher severity level if the worsening threshold $\theta$ is exceeded.
		
		\item \textbf{Care Allocation.}  
		For each patient-caregiver dyad, the model selects a caregiving strategy or service session based on availability, caregiver capacity, and spatial proximity to services.  
		The selected session $s_t = (\lambda, ic, p, sp)$ is executed, consuming caregiver time and updating both effort and walkability indices.
		
		\item \textbf{Functional Reassessment.}  
		Following each session, the patient’s functional condition is re-evaluated.  
		Assistance needs, walkability, and mobility attributes are updated accordingly, influencing the configuration of future sessions.
		
		\item \textbf{Computation of Indices.}  
		Model-level performance indicators-Walkability Index ($WKB$), Caregiver Effort Index ($CEI$), Caregiver Overwhelmed ($CO$), and Hours Not Cared ($HNC$)-are recomputed and stored for analysis.
	\end{enumerate}
	
	This synchronous, stepwise structure enables the simulation of evolving needs, resource utilization, and care dynamics over time.  
	Although discrete-time updating may introduce temporal alignment artifacts, it provides a transparent and computationally tractable framework for studying emergent accessibility patterns and caregiver-patient interactions.  
	For processes requiring finer temporal resolution, such as scheduled appointments or transport coordination, the model allows sub-daily subdivisions into sessions (morning, afternoon, evening), ensuring adequate representation of intra-day variability without introducing unnecessary computational complexity.
	
	Overall, the discrete-time approach provides a robust framework for capturing both micro-level adaptation and macro-level system evolution, balancing interpretability with computational efficiency.
	
\subsection{Design Concepts}

The model’s design follows the ODD-D principles, emphasizing emergence, adaptation, heterogeneity, and stochasticity.  
Agents interact within a spatially explicit environment where accessibility, service reachability, and caregiver burden evolve as a result of decentralized decisions.

\paragraph{Emergence.}  
System-level patterns-such as the spatial distribution of walkability or the clustering of overwhelmed caregivers-arise from micro-level interactions between patients, caregivers, and service locations.  
No global optimization mechanism is imposed; collective outcomes emerge endogenously from agents’ adaptive strategies and spatial constraints.

\paragraph{Adaptation.}  
Patients and informal caregivers adapt their behaviors dynamically in response to environmental and social feedback.  
Patients modify service-seeking decisions based on accessibility, mobility, and aging-stage severity.  
Caregivers reallocate time and effort to balance competing demands across dependents, constrained by mobility and employment attributes.  
This adaptation generates temporal fluctuations in accessibility and effort indices, reflecting realistic behavioral responses to changing local conditions.

\paragraph{Heterogeneity.}  
Both agent classes exhibit diversity in demographic, economic, and mobility attributes, ensuring non-uniform exposure to spatial opportunities and constraints.  
This heterogeneity is a key driver of inequality, influencing emergent accessibility and care distribution patterns.

\paragraph{Stochasticity.}  
Several processes-such as illness onset, aging transitions, or spatial assignment-are probabilistically determined to capture real-world uncertainty.  
Stochastic variation ensures that aggregate patterns are statistically robust rather than deterministic artifacts.

\paragraph{Key Performance Indicators (KPIs).}  
A set of quantitative indicators is recorded to reflect trade-offs among stakeholder priorities.  
These KPIs emerge from agent-based interactions, shaped by evolving preferences and adaptive strategies.  
Formal definitions are provided in Appendix~A.4, while a brief summary follows.

From the \emph{patient} perspective, monitored KPIs include:
\begin{itemize}[nosep]
	\item \textbf{Access distance and modality} - mean distance and type of transport or walking used to reach services;
	\item \textbf{Average abilities} - aggregated functional capacity across aging stages;
	\item \textbf{Average walkability} - the spatial accessibility score reflecting environmental quality and service proximity.
\end{itemize}

From the \emph{informal caregiver} perspective, monitored KPIs include:
\begin{itemize}[nosep]
	\item \textbf{Burden of care and overwhelming status} - cumulative effort compared against the effort threshold;
	\item \textbf{Access time and access distance} - total travel effort per assisted patient;
	\item \textbf{Hours of care not provided} - residual unmet needs due to limited availability or high workload.
\end{itemize}

All indicators are computed cumulatively over sessions and dyads, enabling detailed assessment of spatial, temporal, and behavioral trade-offs in care provision.  
Service utilization can also be derived from the aggregation of available working hours at the service-provider level.

This design supports multi-perspective analysis of care sustainability, bridging micro-level behavioral mechanisms with system-level outcomes that can inform policy evaluation.

	\subsection{Submodels}
Internal components are organized as sub-models, each encapsulating a specific subsystem of behavior and computation.

\subsubsection{Road Network and Graph}

The spatial environment is formalized as a graph $G = (V, E)$ representing the municipal road network derived from high-resolution GIS data.  
Vertices $v_i \in V$ correspond to intersections, buildings, or transport nodes; edges $e_{ij} \in E$ denote directed street segments enriched with semantic attributes.  
Distances between coordinates are computed using the haversine formula, while GIS layers encode features such as elevation, slope, stair presence, and surface type.  
Environmental metadata (e.g., lighting, safety, or pedestrian restrictions) are stored as node attributes influencing mobility cost.  
Buildings and services are spatially embedded within this graph, distinguishing between residential and service structures, each with contextual descriptors such as function, accessibility, and service type.  
Transport nodes (e.g., bus stops) include metadata on capacity and multimodal connectivity.  
Physical barriers-stairs, steep slopes, pedestrian-only areas-annotate nodes with accessibility penalties, allowing agents’ route choice to reflect differentiated terrain effects.  
This transformation from geospatial data to a structured network enables the use of network-analytic computations for evaluating accessibility, walkability, and caregiver burden, thus linking physical space to individual-level constraints.

\subsubsection{ Session Cost Assessment}

Each caregiving or service session quantifies three primary metrics: total distance (km), time (hours), and energy expenditure (kcal).  
For every traversed segment:
\begin{itemize}[nosep]
	\item \textbf{Geodesic distance} is computed via the haversine formula, accounting for Earth curvature.
	\item \textbf{Dynamic speed assignment} adjusts agent speed by mobility mode (walk, vehicle, transit) and context (approach, entry, exit).
	\item \textbf{Travel time} is $t = d / v$, where $d$ is segment distance and $v$ is current speed in km/h.
	\item \textbf{Caloric output} uses metabolic equivalent of task (MET) values: $3.8$ for walking and $2.0$ for vehicular modes, scaled by age and sex.
\end{itemize}

Each agent (and by extension each dyad) has a mobility preference $m_a \in \{\text{private}, \text{public}, \text{walk}, \text{green}\}$ determining session cost.  
An intensity multiplier, based on aging stage and activity type, adjusts the total effort.  
Contextual cost components (traffic, waiting time, stress, or infrastructural barriers) further modify accessibility, capturing factors not evident in spatial distance alone.  
Calibration may rely on empirical surveys or expert elicitation; where unavailable, baseline values validated through internal consistency checks are employed.

\subsubsection{Terrain Features and Indicators}

Each vertex is evaluated against its adjacent nodes to compute slope, fatigue, and derivative indicators affecting perceived walkability.  
Given elevations $z_i, z_j$ and geographic coordinates $(\phi_i, \lambda_i)$ and $(\phi_j, \lambda_j)$, the slope is:
\begin{equation}
s_{ij} = 100 \times \frac{|z_j - z_i|}{\text{haversine}(\phi_i,\lambda_i,\phi_j,\lambda_j)}.
\end{equation}
Fatigue $f_{ij}$ aggregates slope and elevation into a normalized cost surface.  
From $f_{ij}$, three perceptual indices are derived:
\begin{align}
	\text{Difficulty:} &\quad D = 
	\begin{cases}
		0, & f < \tau_1,\\
		(f-\tau_1)^2, & f \ge \tau_1;
	\end{cases} \\
	\text{Safety:} &\quad S = \max(0, 1 - \alpha f); \\
	\text{Pleasantness:} &\quad P = e^{-\beta f}.
\end{align}
Parameters $(\tau_1,\alpha,\beta)$ calibrate thresholds for perceptual change.  
These enriched indicators are embedded in the node-level attribute set and affect routing, fatigue, and accessibility.  
Affective calibration can further adjust these metrics using biosignal data (e.g., galvanic skin response, HRV, muscle activation), regressed against terrain features to better represent anxiety, effort, or discomfort effects observed in elderly mobility studies \citep{bandini2021,bandini2023,gasparini2020}.

\subsubsection{Walkability Index}

For an elder agent $a$ at node $n$, walkability $WKB_{a,n}$ is computed as:
\begin{equation}
WKB_{a,n} = c \times 
\frac{w_s + r + s_f + p_f}{d + \sigma + \delta + \eta},
\end{equation}
where $c$ is a calibration constant (typically $4.8$ for scaling to [0, 50]).  
The components include:  
$w_s$ (sidewalk width, 0-3),  
$r$ (relief score 0-1),  
$s_f$ (safety 0-10),  
$p_f$ (pleasantness 0-10),  
$d$ (slope 0-15),  
$\sigma$ (stairs penalty 0-1),  
$\delta$ (fatigue 1-10),  
and $\eta$ (difficulty).  
Agent-specific adjustment coefficients modify perception with age:
\begin{equation}
\theta_i = \left(\frac{a}{\bar{a}}\right)^k \gamma_i,
\end{equation}
where $\bar{a}$ is the mean age of the cohort, $k$ an exponent ($k=2$ for safety and fatigue, $k=3$ for difficulty), and $\gamma_i$ a scaling constant (0.85 for safety, 1.1 otherwise).  
The walkability index is bounded, monotonic, and Lipschitz-continuous, ensuring smooth comparability across territories.

\subsubsection{Informal Caregiver Efforts Index}

The \textit{Caregiver Effort Index} (CEI) quantifies burden by integrating economic, temporal, and psychosocial dimensions calibrated from structured survey data.  
Each caregiver agent $ic$ has a care capacity $C_{ic}$ determined by income $I_{ic}$, support network $S_{ic}$, and mobility burden $B_{ic}$:
\begin{equation}
C_{ic} = \theta_0 + \theta_1 I_{ic} + \theta_2 S_{ic} - \theta_3 B_{ic}.
\end{equation}
Effort $E_{ic}$ converts required care hours $h_r$ into a normalized index:
\begin{equation}
E_{ic} = \frac{h_r^\alpha}{h_r^\alpha + \beta^\alpha}, \quad \alpha,\beta>0,
\end{equation}
bounded between 0 and 1.  
Perceived performance $P(E)$ follows an inverted-U relation representing burnout:
\begin{equation}
P(E) = \frac{E}{aE^2 + bE + c},
\end{equation}
with $a,b,c>0$ and coefficients modulated by the level of social support $S_{ic}$.  
High support shifts the performance peak rightward, delaying burnout; low support accelerates decline.  
This formulation captures empirical evidence linking excessive caregiving to reduced wellbeing and efficiency.  
All effort and performance scores are stored for post-simulation analysis, enabling comparison of caregiver strain across dyads and spatial contexts.

Both $E$ and $P(E)$ are bounded, continuous, and monotonic in their relevant domains, satisfying stability and scale-invariance conditions for index comparability across municipalities and scenarios.

	\section{Synthetic Population Generation}
	Population synthesis follows a two-stage IPF + imputation pipeline: municipal marginals from ISTAT 8MilaCensus, microdata from AVQ survey, and structured caregiver interviews. IPF aligns socio-demographic distributions; imputation adds care attributes as soft constraints. This hybrid approach preserves plausible joint patterns while avoiding over-fitting.

	\subsection{Concept and Data Foundations}
	
	In population synthesis, \textit{iterative proportional fitting} (IPF) calibrates a sample of individual records--the \textit{seed microdata}--to reproduce known marginal or joint distributions for a target population.  
	The seed microdata represent the initial pool of plausible individuals, households, or dyads, typically drawn from regional or national surveys (e.g., AVQ, ISTAT microdata).  
	Each record includes all variables relevant to the simulation (e.g., age, sex, education, household composition).
	
	Because seed microdata rarely match the small-area marginals of a municipality, IPF iteratively adjusts record weights so that their weighted aggregates conform to the externally supplied target distributions.  
	The method requires that:
	(i) all variables used in constraints exist in the microdata;  
	(ii) categories correspond exactly to those in the marginals; and  
	(iii) every combination of constrained categories appears at least once in the microdata.
	
	\textit{Marginals} represent the distributions of individual variables (e.g., age or sex), typically obtained from census or administrative data.  
	\textit{Joint distributions} capture dependencies between variables (e.g., age-sex-education), ensuring that co-occurrence patterns are preserved.  
	Using joint constraints prevents unrealistic independence among attributes, producing a more coherent synthetic population.
	
	After convergence, each record obtains a weight $w_i$ proportional to its representation in the target population.  
	Sampling with replacement according to these weights yields the synthetic population of the desired size (e.g., 500 elderly individuals per municipality).  
	Records with higher $w_i$ are selected more frequently, ensuring that local demographic proportions align with all constraints even when seed data are regional.

	\subsection{Handling Missing or Under-Sampled Attributes}

	When microdata lack attributes required by the constraints or when soft priors are available, three approaches can be adopted:
	
	\begin{enumerate}[label=(\alph*)]
		\item \textbf{IPF with post-imputation:}  
		Apply IPF on available variables, then impute missing attributes assuming conditional independence given balanced covariates.
		\item \textbf{Joint probabilistic modeling:}  
		Use Bayesian networks or hierarchical models to represent dependencies among observed and missing variables; this approach is more flexible but computationally intensive.
		\item \textbf{Synthetic reconstruction:}  
		Generate agents stochastically from assumed distributions when no detailed microdata exist, trading empirical realism for generality \citep{chapuis2022}.
	\end{enumerate}

	\subsection{Algorithmic Overview}
	
	The Two-Step IPF with Marginal and Joint Constraints proceeds as follows:
	
	\begin{enumerate}[label=\arabic*.]
		\item \textbf{Initialization:}  
		Each record $i$ in the seed microdata is assigned a unit weight, $w_i = 1$.
		
		\item \textbf{Iterative Weight Update:}
		\begin{itemize}[nosep]
			\item For each marginal variable $V$ and category $c$:
			\begin{equation}
w_i^{(k+1)} = w_i^{(k)} \times \frac{\text{target\_prop}(V=c)}{\text{weighted\_sum}(V=c)}.
\end{equation}
			\item For each joint combination $(V_1,V_2,\ldots,V_j)$ and category tuple $(c_1,\ldots,c_j)$:
			\begin{equation}
w_i^{(k+1)} = w_i^{(k)} \times \frac{\text{target\_prop}(c_1,\ldots,c_j)}{\text{weighted\_sum}(c_1,\ldots,c_j)}.
\end{equation}
		\end{itemize}
		
		\item \textbf{Convergence Check:}  
		Iterations continue until the maximum absolute weight change satisfies  
		$\max_i |w_i^{(k+1)} - w_i^{(k)}| < \varepsilon$,  
		where $\varepsilon$ is a user-defined tolerance (e.g., $10^{-6}$).
		
		\item \textbf{Imputation (optional):}  
		For under-specified attributes, marginal probabilities are estimated and values are stochastically assigned via random draws  
		$x_i \sim \text{Multinomial}(p_1,\ldots,p_k)$,  
		where $p_j$ represents the proportion for category $j$.
	\end{enumerate}
	
	This iterative process ensures that the weighted microdata replicate both marginal and joint constraints to numerical precision.  
	Weighted sampling then produces a synthetic population consistent with local demographic and socioeconomic patterns, suitable for agent-based simulation and spatial policy evaluation.

	\section{Sensitivity Analysis}
	Global sensitivity combines Morris screening and Sobol-Saltelli variance decomposition on key parameters (caregiver income, pension income, mobility adoption, decay $\beta$). Results show interaction-driven behaviour for walkability and dominant financial effects for hours not cared.
	
	\subsection{Sensitivity Analysis of Structural Parameters}
	
	To assess the influence of structural parameters on the model’s key performance indicators (KPIs), two complementary global sensitivity methods were applied:  
	(i) the \textit{Morris elementary-effects} method for qualitative screening of first-order effects and interaction potential, and  
	(ii) the \textit{Sobol-Saltelli variance decomposition} for quantitative estimation of main and total-order contributions.  
	Both analyses were performed on two experimental batches of 40 runs each (corresponding to Scenarios~1 and~2), examining three key parameters: baseline income ($X_1$), pensions ($X_2$), and mobility adoption ($X_3$).
	
	Across experiments, results converge on a consistent pattern: financial and mobility factors jointly shape the model’s emergent accessibility and caregiving dynamics, with varying relative importance across KPIs.
	
	\paragraph{KPI1 - Caregiver Effort Index.}  
	Both Morris and Sobol analyses identify all three parameters as influential, with pensions ($X_2$) and income ($X_1$) generally exerting stronger effects than mobility adoption ($X_3$).  
	Variability in the estimates suggests non-linear interactions, but the consistent finding is that financial resources-wages and pensions-remain the most robust predictors of caregiving effort.  
	Sustainability arises from the combined influence of financial and mobility capacities, echoing empirical evidence that caregiving intensity is linked to both economic stress and accessibility \citep{colombo2011}.
	
	\paragraph{KPI2 - Number of Caregivers in Burnout.}  
	Both methods highlight substantial contributions from pensions ($X_2$) and mobility adoption ($X_3$), with income ($X_1$) playing a secondary role.  
	The balance between $X_2$ and $X_3$ varies across runs, but the recurrent pattern indicates that burnout risk cannot be explained by income alone.  
	Instead, it reflects the interplay between financial security and access to mobility options, illustrating how low-income households remain particularly vulnerable under constrained mobility conditions.
	
	\paragraph{KPI3 - Walkability Index.}  
	This indicator exhibits the clearest evidence of interaction-driven behavior.  
	Morris indices suggest moderate main effects for all parameters, while Sobol results show weak first-order but consistently high total-order indices, confirming strong interaction effects.  
	Walkability therefore emerges not from any single driver but from the joint influence of income, pensions, and mobility adoption.  
	This robustness across scenarios underscores that accessibility improvements depend on multidimensional resource integration, consistent with evidence linking income, services, and transport availability in rural and aging areas \citep{shergold2012}.
	
	\paragraph{KPI4 - Hours Yet to Care.}  
	Results are more stable: both Morris and Sobol analyses indicate that pensions ($X_2$) and income ($X_1$) dominate explanatory power, with mobility adoption ($X_3$) contributing marginally.  
	The relative ordering between $X_1$ and $X_2$ fluctuates across runs, but the underlying structure remains constant-unmet care demand is driven primarily by financial endowments, with mobility effects secondary.
	
	Overall, the sensitivity analyses demonstrate that financial resources and mobility jointly underpin care sustainability, while interaction effects play a crucial role in shaping accessibility outcomes.  
	These findings reinforce the need for policies that address both economic and spatial dimensions of vulnerability rather than treating them in isolation.

\section{Initialization, Warm-Up, and Run Design}
The model is initialized with a synthetic population and a predefined set of structural parameters controlling demographic composition, behavioral attributes, and mobility capacities. Such parameter values are derived from a combination of empirical data from Italian transport and mobility datasets \citep{istat2022mobility,regionelombardia2020mobility,caselli2021,rossetti2020}, literature-based values such as average walking speed and service density follow gerontological and accessibility studies \citep{bohannon1997,studenski2011,colombo2011,vale2016}, and internal calibration runs. Table~\ref{tab:init-params} summarizes the basis for selecting these input parameters at initialization, distinguishing between empirical, statistical, and literature-derived sources. This initialization ensures that simulated agents represent realistic demographic and socioeconomic conditions of Italian inner areas while maintaining generalizability across municipalities.

The simulation starts from an empty state, with agents and services progressively instantiated according to the defined parameters. 
A warm-up period of approximately 30 iterations is determined using the Schruben-Singh-Tierney procedure with batch means to minimize autocorrelation and eliminate initialization bias. 
Each scenario is then executed for an extended duration with 40 independent replications, supporting robust estimation of confidence intervals for all key performance indicators (KPIs).

		\begin{table}[t]
	\centering
	\small
	\setlength{\tabcolsep}{8pt}
	\renewcommand{\arraystretch}{1.15}
		\caption{Basis for the selection of input parameters used at initialization (Section~5).}
	\label{tab:init-params}
	\begin{tabular}{llll}
		\toprule
		\textbf{Attribute} & \textbf{Description} & \textbf{Value} & \textbf{Basis (Source)} \\
		\midrule
		\texttt{wk\_speed}        & Walking speed (km/h)                 & 4    & Literature \\
		\texttt{pr\_speed}        & Private mobility speed (km/h)         & 40   & Empirical \\
		\texttt{pu\_speed}        & Public mobility speed  (km/h)         & 30   & Empirical \\
		\texttt{gr\_speed}        & Alternative mobility speed (km/h)    & 10   & Empirical \\
		\texttt{gtperc}           & Walking willingness (fraction)             & 0.3  & Empirical \\
		\texttt{radius\_preference} & Walkability radius (world units)           & 2.5  & Empirical \\
		\texttt{age\_ref}         & Elderness threshold (years)          & 65   & ISTAT \\
		\texttt{age\_work}        & Working age (years)                    & 15   & ISTAT \\
		\texttt{age\_gold}        & Older elders (years)                   & 75   & ISTAT \\
		\texttt{efforts\_threshold} & Threshold for efforts        & 2.5  & Index/Survey \\
		\texttt{base\_income}     & Base income (monthly, EUR)          & 1200 & ISTAT \\
		\texttt{base\_income\_ret}& Pension income (monthly, EUR)       & 700  & ISTAT \\
		\texttt{hrs\_ass\_max}    & Hours of assistance (max)          & 12   & Index \\
		\texttt{n\_facilities}    & Services considered            & 5    & Literature \\
		\bottomrule
	\end{tabular}

	\footnotesize{\textit{Note:} Units follow the main text; speeds are nominal scenario values.}
\end{table}

	\subsection{Warm-up Duration}
	
	To determine an appropriate warm-up period, the baseline scenario was simulated over six months. Figure~6 illustrates the stabilization of agent populations, average access times, caregiver burden, and walkability indices. Visual inspection suggests convergence between iterations~20 and~40. A formal assessment of initialization bias was conducted using the Schruben-Singh-Tierney test \citep{schruben1983} at a two-tailed 5\% significance level, combined with a batch-means method using five batches \citep{kleijnen1984,kleijnen2007}. Results indicated that a minimum of 30 iterations was required for stable outputs across all indicators. The selected warm-up period thus ensures statistical stationarity of mean estimates before data collection. Further methodological details are provided in Appendix~A.7.
	
	\subsubsection{Initialization Bias and Batch Estimation}
	
	To mitigate initialization bias, the model’s start-up transients were examined using both visual and statistical diagnostics. The procedure comprised three steps:
	
	\begin{enumerate}[label=(\arabic*)]
		\item \textbf{Stability inspection:} Long simulations (10$\times$ expected duration) were performed for key outputs-average travel time to ambulatory services, caregiver effort index, and proportion of population receiving adequate care. Moving averages (10-50 step window) were used to identify when trajectories stabilized.
		\item \textbf{Quantitative stability test:} Rolling mean and the corresponding normalized mean squared error (HMSE) were computed. When HMSE fell below 0.02, the system was deemed stable. The maximum stabilization point across variables plus a 100-500~step margin was used as warm-up duration.
		\item \textbf{Batch size determination:} Five preliminary replications were conducted using fixed parameters (post-warm-up). For each output, sample mean and standard deviation ($\sigma$) were used to determine batch length ensuring a 95\% confidence interval with a $\pm$5\% margin of error ($E = 1.96\, \sigma / \sqrt{n}$). The maximum batch size across variables was adopted.
	\end{enumerate}
	
	These procedures collectively ensure that transient initialization effects are excluded and that steady-state means are estimated with consistent precision across all monitored variables.

	\section{Stochasticity and Random Number Generation}
	The model uses NetLogo’s MT19937 (Mersenne Twister) RNG with per-agent sub-streams derived from the global seed and unique agent IDs \cite{netlogo}. Replications use distinct seeds. Randomness influences key processes governing agent initialization, health evolution, behavioral preferences, and event timing. Stochastic elements ensure heterogeneity across dyads and support realistic variability in outcomes, rather than deterministic trajectories.
	
	\subsection{Selection of Dyads and Frequency of Care Needs}
	
	The daily distribution of care needs follows a Gaussian process that models temporal fluctuations in demand across three periods-morning, afternoon, and evening.  
	For each time block $b$, the probability that an elder requires care at hour $t$ is given by a normalized Normal density:
	\begin{equation}
P_b(t) = \frac{f(t; \mu_b, \sigma_b)}{\int_{t_b^-}^{t_b^+} f(u; \mu_b, \sigma_b)\, du},
\end{equation}
	where $f(t; \mu_b, \sigma_b)$ is the Gaussian PDF centered on peak hour $\mu_b$ with standard deviation $\sigma_b$.  
	The expected number of elders requesting care at time $t$ is $N_b \cdot P_b(t)$, where $N_b$ is the configured proportion of elders needing care in block $b$.  
	This modular Gaussian scheme reproduces diurnal variations in care activities-morning peaks for routine assistance, afternoon for physical fatigue, and evening for social or emotional support-and can be replaced by a Poisson process when event arrivals are discrete or memoryless.
	
	\subsection{Aging Progress}
	
	Transitions between aging stages and mortality are modeled as stochastic events.  
	Aging progression is perturbed by uniform noise $\epsilon \sim U[-a,a]$, where $a$ is a stage-specific variability factor.  
	Mortality follows a Bernoulli process with success probability equal to the stage-dependent annual death rate $p_s$, evaluated at each simulated year.  
	Alternative formulations allow normally distributed offsets or empirically fitted survival probabilities to capture heterogeneity in health trajectories.
	
	\subsection{Walking Distance Preference}
	
	Each agent’s walking preference (in world units) is sampled as:
	\begin{equation}
r \sim U[1.5, 3.5],
\end{equation}
	representing the maximum Euclidean distance an individual is typically willing to walk to reach a service.  
	This distribution reflects differences in physical ability and spatial context.  
	When desired, a truncated normal distribution $r \sim \mathcal{N}(\mu=2.5, \sigma=0.5)$ bounded to $[1.5,3.5]$ may be used to simulate smoother variation.
	
	\subsection{Hours of Assistance}
	
	Assistance demand increases with aging stage. For stage $i \in \{1,\dots,4\}$, required hours are sampled from:
	\begin{equation}
H_i \sim U(\mu_i - \Delta_i,\, \mu_i + \Delta_i),
\end{equation}
	where $\mu_i$ is the stage-specific mean and $\Delta_i$ defines bounded heterogeneity.  
	This captures variability due to health conditions, autonomy, and environment.
	
	\subsection{Onset and Selection of Assistance Sessions}
	
	At each decision step, agents select an assistance session $s_j \in S$ using one of three stochastic modes:
	\begin{enumerate}[label=(\roman*)]
		\item \textbf{Uniform:} all sessions equally likely ($P(s_j)=1/n$);
		\item \textbf{Empirical:} probabilities derived from observed or interview-based frequencies, sampled via inverse transform on cumulative probabilities;
		\item \textbf{Weighted:} user-specified weights converted to normalized probabilities.
	\end{enumerate}
	These options allow behavioral experiments ranging from random exploration to empirically grounded, preference-driven decision-making.
	
	\subsection{Income Variability}
	
	Disposable income evolves as a stochastic process combining deterministic and random components:
	\begin{equation}
I(t) = I_0 + \beta t + A \sin(2\pi f t) - \delta,
\end{equation}
	where $I_0$ is base income, $\beta$ the trend, and $A$ and $f$ capture seasonal variation.  
	A random deduction $\delta \sim U[0, \rho I_0]$ introduces exogenous shocks (e.g., expenses or health costs).  
	This reflects income volatility typical of aging and caregiver populations.
	
	\subsection{Service Duration}
	
	Service duration times are stochastic and, where available, drawn from empirical distributions fitted to log-normal models by service type:
	\begin{equation}
T_s \sim \text{LogNormal}(\mu_s, \sigma_s).
\end{equation}
	In the absence of data, durations are randomly perturbed by $\Delta t \sim U(0,30)$ minutes, allowing the representation of incidental delays and unpredictable events.

	\begin{table}[t]
		\centering
		\setlength{\tabcolsep}{8pt}\renewcommand{\arraystretch}{1.1}
		\begin{tabular}{l l}
			\toprule
			\textbf{Aspect} & \textbf{Distribution} \\
			\midrule
			Dyad selection per tick & Gaussian / rule-based \\
			Ageing process (incl.\ death) & Uniform with Bernoulli trial \\
			Walkable range preferences & Uniform or Gaussian \\
			Daily hours of assistance & Uniform, stage-specific \\
			Unmet attempts / availability & Rule + schedule constraint \\
			Unexpected expenses / time & Uniform, time-dependent \\
			\bottomrule
		\end{tabular}
		\caption{Stochastic components and their implemented distributions.}
		\label{tab:stochastic_components}
	\end{table}

	\subsection{Reproducibility}
	
	To ensure decorrelation and reproducibility, a hashing-based seeding strategy was developed. Each agent’s seed is computed as:
	\begin{equation}
\texttt{agent-seed} = \texttt{mix32}\left(\texttt{global-seed} \oplus (\,\texttt{who} \times 0x9E3779B9\,)\right)
\end{equation}
	where \texttt{mix32} is a compact XOR/shift finalizer applying avalanche scrambling within the signed 32-bit domain. This approach avoids 64-bit arithmetic (unsupported in NetLogo), guarantees reproducible per-agent streams, and prevents correlation between identifiers.
	
	Diagnostic tests confirmed the robustness of this seeding scheme: $k$-distribution chi-square tests showed no systematic deviation from uniformity, and between-agent correlations remained within theoretical bounds. The method draws upon established principles of multiplicative hashing \citep{knuth1998} and xorshift-based randomization \citep{marsaglia2003,chi2017}, ensuring independence, reproducibility, and computational efficiency for large-scale agent-based simulations.